\documentclass[a4paper,11pt]{article}
\pdfoutput=1 % if your are submitting a pdflatex (i.e. if you have
\usepackage{jcappub} % for details on the use of the package, please
\usepackage[T1]{fontenc} % if needed

\usepackage{graphicx}
\usepackage{dcolumn}
\usepackage{bm}
\usepackage{hyperref}
\usepackage{xcolor}
\usepackage{ulem}
\usepackage{eucal}
\usepackage{subcaption}
\allowdisplaybreaks
\renewcommand{\thefigure}{\thesection.\arabic{figure}}
\usepackage{comment}

\newcommand{\firstphi}{\phi^{(1)}}
\newcommand{\firstpsi}{\psi^{(1)}}

\newcommand{\tensorh}{h^{i(2)}_{j}}

\newcommand{\dfirstphi}{\phi^{(1)'}}
\newcommand{\dfirstpsi }{\psi^{(1)'}}

\newcommand{\ddfirstphi}{\phi^{(1)''}}
\newcommand{\ddfirstpsi}{\psi^{(1)''}}

\newcommand{\ddtensorh}{h^{i(2)''}_{j}}
\newcommand{\dtensorh}{h^{i(2)'}_{j}}

\newcommand{\dddfirstphi}{\phi^{(1)'''}}
\newcommand{\dddfirstpsi}{\psi^{(1)'''}}

\newcommand{\ddddfirstpsi}{\psi^{(1)''''}}

\newcommand{\firstphigr}{\phi^{(1)}_{\text{GR}}}
\newcommand{\dfirstphigr}{\phi^{(1)'}_{\text{GR}}}
\newcommand{\ddfirstphigr}{\phi^{(1)''}_{\text{GR}}}
\newcommand{\dddfirstphigr}{\phi^{(1)'''}_{\text{GR}}}
\newcommand{\ddddfirstphigr}{\phi^{(1)''''}_{\text{GR}}}

\usepackage[toc,page]{appendix}
\usepackage{bm}
\newcommand\numberthis{\addtocounter{equation}{1}\tag{\theequation}}
\usepackage[capitalise]{cleveref}

\usepackage{natbib}
\title{\boldmath Scalar-Induced Gravitational Waves in Modified Gravity}

\author[a,b]{Anjali Abirami Kugarajh,}
\author[c,i]{Marisol Traforetti,}
\author[a,b]{Andrea Maselli,}
\author[c,d,f,a]{Sabino Matarrese,}
\author[g,h,c]{Angelo Ricciardone}

\affiliation[a]{Gran Sasso Science Institute,
Viale F. Crispi 7, I-67100 L’Aquila, Italy.}
\affiliation[b]{INFN, Laboratori Nazionali del Gran Sasso, I-67100 Assergi, Italy.}
\affiliation[c]{Dipartimento di Fisica e Astronomia “Galileo Galilei”, Universit\`a degli Studi di Padova, Via Marzolo 8, I-35131, Padova, Italy.}
\affiliation[d]{INFN, Sezione di Padova, Via Marzolo 8, I-35131 Padova, Italy.}
\affiliation[f]{INAF, Osservatorio Astronomico di Padova, Vicolo dell’Osservatorio 5, I-35122 Padova, Italy.}
\affiliation[g]{Dipartimento di Fisica “Enrico Fermi”, Università di Pisa, Largo Bruno Pontecorvo 3, Pisa I-56127, Italy.}
\affiliation[h]{INFN, Sezione di Pisa, Pisa I-56127, Italy.}
\affiliation[i]{Institut de Ciències del Cosmos (ICC), Universitat de Barcelona, Martí i Franquès 1, ES-08028, Barcelona, Spain.}

\emailAdd{anjali.kugarajh@gssi.it}
\emailAdd{marisol.traforetti@icc.ub.edu}
\emailAdd{andrea.maselli@gssi.it}
\emailAdd{sabino.matarrese@pd.infn.it}
\emailAdd{angelo.ricciardone@unipi.it}

\abstract{Scalar-Induced Gravitational Waves (SIGWs) — second-order tensor modes sourced by first-order scalar fluctuations in General Relativity (GR)— are expected to contribute to the Stochastic Gravitational Wave Background (SGWB) potentially detectable by current and future gravitational wave interferometers. In the framework of GR, this SGWB represents an unavoidable contribution to the gravitational wave spectrum. In this paper, we go beyond GR and we investigate the behavior of SIGWs in $f(R)$ gravity. We explore how the SIGW spectrum is influenced across a broad range of frequencies, from the nano-Hz regime, where the Pulsar Timing Array (PTA) operates, through the milli-Hz band probed by the space-based LISA detector, up to the kilo-Hz frequency range, where the ground-based LIGO/Virgo/KAGRA network is currently operational. Our results indicate that the beyond-GR correction leaves an observational imprint, mainly in the low-frequency part of the spectrum, giving the possibility to use SIGW to constrain GR on scales on which we have limited information.}

\begin{document}

\maketitle
\flushbottom

\section{Introduction}
\label{sec:introduction}

Einstein's theory of General Relativity (GR) is seen as a basis for the description of our universe. One of the main predictions of Einstein's theory is the existence of gravitational waves (GWs). Direct detection of GWs from mergers of binary black holes by the LIGO-Virgo-KAGRA collaboration has not only provided stringent tests of GR but it has additionally given a new approach to probe fundamental physics. Inflationary cosmology predicts the generation of quantum vacuum fluctuations of metric perturbations, where we observe scalar modes as density fluctuations and tensor modes as gravitational waves \cite{Guth:1982ec, Wands:2008tv}, which weakly interact with matter \cite{Bartolo:2019oiq, Bartolo:2019yeu}. Primordial GWs are assumed to come in the form of a stochastic background (SGWB), due to the quantum nature of fluctuations. Various mechanisms active in the early Universe can source GWs \cite{Maggiore:2007ulw, Maggiore:2018sht, Caprini:2018mtu, Guzzetti:2016mkm}. One contribution arises from ``scalar-induced'' GWs (SIGWs), which result from second-order effects and are sourced by first-order scalar fluctuations \cite{10.1143/PTP.37.831, PhysRevD.47.1311, Matarrese:1993zf, Pantano:1995wvf, Matarrese:1997ay,Carbone:2004iv, Ananda:2006af, Baumann:2007zm, Domenech:2021ztg}. At linear order, the tensor modes propagate independently of the scalar modes. However, when studying the metric and the stress-energy tensor at second order, it appears that the evolution of second-order tensor modes is sourced by the coupling of first-order fluctuations. This is an outcome of the non-linearity of Einstein's equations. Additionally, since SIGWs are also sensitive to the underlying statistics of scalar fluctuations \cite{Nakama:2016gzw,Perna:2024ehx,Li:2023xtl,Li:2024zwx,Atal:2021jyo,Unal:2018yaa,Yuan:2023ofl} they can be used to probe primordial non-Gaussianity (see \cite{Perna:2024ehx, LISACosmologyWorkingGroup:2025vdz} for a recent forecast for the LISA detector).  SIGWs are mainly sourced by the coupling of scalar modes, but more recently there have been studies about scalar-tensor couplings \cite{Bari:2023rcw} and first-order tensor perturbations \cite{Picard:2023sbz}. In turn, tensor modes have been shown to produce density perturbations \cite{10.1143/PTP.45.1747, Matarrese:1997ay, Bari:2021xvf, Bari:2022grh,Bertacca:2024zfb}. Also anisotropies from SIGW are a powerful tool to distinguish them from other cosmological background sources \cite{Bartolo:2019zvb,LISACosmologyWorkingGroup:2022kbp,ValbusaDallArmi:2023nqn, ValbusaDallArmi:2024hwm}.

Previous works investigated the primordial spectrum of SIGWs in the radiation-dominated era, then extended to the matter-dominated era \cite{Ananda:2006af,Baumann:2007zm}.
In all these studies, the analysis of the GW spectrum is carried out under the assumption that the anisotropic stress is negligible. If this assumption is not made at the beginning, a coupling between first-order scalar mode and to first-order anisotropic stress arises 
as a result of the second-order energy-momentum tensor. For 
this reason, if there is a 
mismatch between the two scalar-modes, this coupling appears in  the source term and possibly in the spectrum. Such a non-zero difference between the scalar potentials is known as {\it gravitational-slip} and it is typically studied in the Modified Gravity (MG) context \cite{Tsujikawa:2007gd,Caldwell:2007cw,Pogosian:2010tj,LIGOScientific:2018dkp,Capozziello:2011et,Clifton:2011jh}. In this case, the effect arises from the modification in the geometry of the trace-less part of the first-order spatial component of the field equation. 

SIGWs are typically studied within the framework of GR, where they already have several intriguing implications. However, recent studies have explored SIGWs in modified gravity theories, such as Horndeski gravity \cite{Domenech:2024drm}, focusing on an early-universe phase dominated by a scalar field, and in $f(R)$ gravity \cite{Zhou:2024doz}. In particular, in the latter, calculations have been performed for the Hu-Sawicki (HS) model~\cite{PhysRevD.76.064004}, with $f(R) \propto R + c R^n$ with $n = 1$. The resulting spectrum has been compared with current observational data from pulsar timing arrays (PTAs).
 
The same large primordial fluctuations that generate SIGWs can also lead to the formation of primordial black holes (PBHs) when they re-enter the horizon and collapse \cite{Zeldovich:1967lct,1974MNRAS.168..399C, Hawking:1971ei,Carr:1975qj,Chapline:1975ojl,Iovino:2024sgs}. As a consequence, SIGWs are an unavoidable byproduct of PBH formation, arising naturally through second-order perturbation effects. This fundamental connection provides a powerful avenue for using SIGWs to place constraints on the abundance of PBHs \cite{Saito:2008jc, Bugaev:2010bb, Nakama:2016gzw,Bartolo:2018evs,Inomata:2018epa}.  Beyond GR, alternative theories of gravity may further influence the interplay between SIGWs and PBHs. For instance, in the context of $f(R)$ gravity with a quadratic correction, $f(R) \propto R^2$, it has been shown that the SGWB induced by PBHs can serve as an independent probe of modifications to gravity \cite{Papanikolaou:2021uhe}. Specifically, this approach enables the derivation of an upper bound on the PBH abundance in modified gravity, which is more stringent than the corresponding constraint in GR. This suggests that the presence of $f(R)$ corrections may impose tighter restrictions on the number density of PBHs, potentially leading to a significant suppression of their formation.  

SIGWs are a benchmark signal for a wide range of current and next-generation detectors, both ground-based and space-based. These include PTA \cite{NANOGrav:2023gor,EPTA:2023fyk,Reardon:2023gzh,Xu:2023wog,Figueroa:2023zhu}, the Laser Interferometer Space Antenna (LISA) \cite{LISACosmologyWorkingGroup:2022jok, Caprini:2019pxz,Kume:2024sbu,LISACosmologyWorkingGroup:2024hsc} and the ground-based LIGO-Virgo-KAGRA (LVK) and Einstein Telescope (ET) \cite{ET:2019dnz, Branchesi:2023mws,Jiang:2024aju,Romero-Rodriguez:2024ldc,Caporali:2025mum}.  Several analysis and forecasts have been performed to understand the ability of such detectors to measure the SIGW signal (see e.g. \cite{LISACosmologyWorkingGroup:2025vdz} for recent analysis). 

In this paper, we investigate the spectrum of SIGWs in a modified gravity framework and compare it to the standard GR prediction. In particular, we focus on an $f(R) = R + \alpha R^2$ gravity model \cite{DeFelice:2010aj}. We derive the equations of motion for first-order cosmological perturbations and second-order SIGWs within this modified gravity scenario. Furthermore, we present the general form of the equations governing second-order SIGWs and their formal solutions. To characterize the impact of modified gravity on SIGWs, we derive the relevant kernel function and perform an averaging procedure. We analyze how the beyond-GR source term affects the spectral energy density of SIGWs. Our study assumes a log-normal scalar power spectrum, where primordial fluctuations exhibit an enhanced amplitude at small scales compared to those constrained by cosmic microwave background (CMB) observations. We numerically compute the SIGW spectrum for both GR and beyond-GR contributions, exploring a range of coupling constants and frequencies relevant to current and future GW detectors. This analysis highlights the potential of SIGWs as a novel multi-band observational tool to probe deviations from GR and constrain alternative gravity theories.

The paper is composed as follows. In section \ref{sec:second_order_gr}, cosmological perturbation theory up to second-order is studied in order to formulate the spectral energy density of the SIGWs. We study the behavior of first-order scalar modes and the equations of motion of the second-order tensor modes, evaluated in the radiation-dominated era. We compute the power spectrum through integrals of oscillating functions and primordial curvature perturbations. In section \ref{sec:second_order_mg}, we extend these results using the same methodology in the context of the $f(R)$ gravity theory and find in section \ref{sec:power_spectrum_modified} the equivalent power spectrum of scalar-induced GWs, taking into account the modification. In section \ref{sec:spectral_energy_density}, using the log-normal primordial scalar power spectrum, we compute the analytical and numerical result from standard GR and $f(R)$ gravity. We compare the results and discuss the implications in section \ref{sec:conclusion}.

Unless specified otherwise, hereafter we use natural units such that $c = \hbar = 1$ and the Planck mass $M_{p}^{-2} = 8\pi G = \kappa^{2}$. Moreover, the prime is used for differentiation with respect to the conformal time $\tau$, with $a(\tau) d\tau = dt$. We denote four-dimensional and spatial indices with Greek and Roman letters, respectively.

\section{\label{sec:second_order_gr}Second-order scalar-induced 
gravitational waves in General Relativity}

In this section, we describe the formalism used to derive the source term of scalar-induced GWs and the corresponding power spectrum. We first introduce a generic scheme to study second-order perturbations of the metric and of the stress-energy tensor, and then compute the equation of motion of scalar-induced GWs. We refer the reader to \cite{Matarrese:1997ay, Ananda:2006af, Baumann:2007zm, Malik:2008im, Kohri:2018awv} for further technical details.

\subsection{\label{subsec:definitions}Definitions and setup}

We consider perturbations around a flat Friedmann-Lemaitre-Robertson-Walker (FLRW) background up to the second-order \cite{Malik:2008im,Matarrese:1997ay} and we split the metric components in terms of scalar, vector and tensor quantities \cite{Bardeen:1980kt}, $\phi^{(r)}$, $\psi^{(r)}$, $w_{i}^{(r)}$ and $h_{ij}^{(r)}$, where the apex $r$ indicates the order of the perturbation.
The perturbed metric in the Poisson gauge reads
\begin{align*}
\label{perturbed_metric}
    ds^{2}  = & a^{2}(\tau)\left\{-\left(1 + 2\firstphi + \phi^{(2)}\right)d\tau^{2} + w_{i}^{(2)}d\tau dx^{i} \right. \\
    & \left. + \left[\left(1 - 2\firstpsi - \psi^{(2)}\right)\delta_{ij} + \dfrac{1}{2}h_{ij}^{(2)}dx^{i}dx^{j}\right]\right\}\ , \numberthis
\end{align*}
where first-order tensor modes are ignored. In addition, we do not include first-order vectors since we know that they have decreasing amplitudes and they are not generated by standard mechanisms \cite{Bartolo:2004if}. Unlike first-order tensor modes, second-order perturbations are gauge dependent, being sourced by the -gauge-dependent- first-order scalar modes \cite{Tomikawa:2019tvi, Inomata:2019yww, DeLuca:2019ufz, Yuan:2019fwv}. A review paper exploring the study of SIGWs across various gauges is currently a work in progress \cite{Anjali2025}.
In General Relativity, the evolution of cosmological perturbations is obtained from the perturbed Einstein’s field equations, 
\begin{equation}
\label{einstein_energy_momenetum_tensor}
    G^{\mu}{_{\nu}}  = R^{\mu}{_{\nu}} - \frac{1}{2} g^{\mu}{_{\nu}}R=\kappa^{2}T^{\mu}{_{\nu}}\ , \numberthis
\end{equation}
where $R$ and $R_{\mu\nu}$ are the Ricci scalar and Ricci tensor, $G_{\mu\nu}$ is the Einstein tensor and $T_{\mu\nu}$ is the energy-momentum tensor.
We consider the energy content of the Universe to be described by a fluid with energy-momentum tensor
\begin{equation}
    T^{\mu}{_{\nu}}  = (\rho + P)u^{\mu}u_{\nu} + P\delta^{\mu}{_{\nu}} + \pi^{\mu}{_{\nu}}
\end{equation}
where $\rho$ and $P$ are the energy density and the isotropic pressure of the fluid, respectively, and $u_\mu$ is the four-velocity of the fluid elements normalized to $u_\mu u^\mu=-1$. $\pi_{\mu\nu}$ is the anisotropic stress tensor and it is subject to the constraints $\pi^\mu_\mu=0$ and $\pi_{\mu\nu}u^\nu = 0$.
Equations~\eqref{einstein_energy_momenetum_tensor} are decomposed at different order in the perturbation.

\subsection{\label{subsec:evolution_equation} GW evolution equation}

The spatial second-order Einstein's field equation is found as
\begin{equation}
G^{(2)i}_{j} = \kappa^{2}T^{(2)i}_{j} \,,  \label{eq: second-order EFE}
\end{equation}
where
\begin{align*}
\label{second_order_spatial_einstein_tensor}
    G^{(2)i}_{j} &= a^{-2}\left[\frac{1}{4}h^{i(2)''}_{j} + \frac{1}{2}\frac{a'}{a}h^{i(2)'}_{j} - \frac{1}{4}\nabla^{2}h^{i(2)}_{j} \right.\\
    & \left. + \partial^{i}\firstphi\partial_{j}\firstphi + 2\firstphi\partial^{i}\partial_{j}\firstphi - 2\firstpsi\partial^{i}\partial_{j}\firstphi - \partial_{j}\firstphi\partial^{i}\firstpsi - \partial^{i}\firstphi\partial_{j}\firstpsi \right. \\
    & \left. + 3\partial^{i}\firstpsi\partial_{j}\firstpsi + 4\firstpsi\partial^{i}\partial_{j}\firstpsi + \frac{1}{2}\left(-\partial^{i}\partial_{j}\phi^{(2)} + \partial^{i}\partial_{j}\psi^{(2)}\right) \right. \\
    & \left. + \frac{1}{2}\frac{a'}{a}\left(-\partial^{i}w_{j}^{(2)} + \partial_{j}w^{i(2)}\right) - \frac{1}{4}\left(-\partial^{i}w_{j}^{(2)'} + \partial_{j}w^{i(2)'}\right) \right. \\
    & \left. + \left(\frac{1}{2}\nabla^{2}\phi^{(2)}+\left[2\frac{a''}{a}-\left(\frac{a'}{a}\right)^{2}\right]\phi^{(2)} + \frac{a'}{a}\phi^{(2)'}\right. \right. \\
    & \left. \left.-\frac{1}{2}\nabla^{2}\psi^{(2)}+\psi^{(2)''} + 2\frac{a'}{a}\psi^{(2)}  + \left[4\left(\frac{a'}{a}\right)^{2} - 8\frac{a''}{a}\right](\firstphi)^{2} \right. \right. \\
    & \left. \left. -8\frac{a'}{a}\firstphi\dfirstphi - \partial_{k}\firstphi\partial^{k}\firstphi - 2\firstphi\nabla^{2}\firstphi  \right. \right. \\
    & \left. \left. - 4\firstphi\ddfirstpsi - 2\dfirstphi\dfirstpsi  - 8\frac{a'}{a}\firstphi\dfirstpsi \right. \right. \\
    & \left. \left. -2\partial_{k}\firstpsi\partial^{k}\firstpsi - 4\firstpsi\nabla^{2}\firstpsi + (\dfirstpsi)^{2} \right. \right. \\
    & \left. \left. + 8\frac{a'}{a}\firstpsi\dfirstpsi + 4\firstpsi\ddfirstpsi + 2\firstpsi\nabla^{2}\firstphi\right)\delta^{i}_{j}\right]\ , \numberthis
\end{align*}
and \footnote{Notice that our derivation of the energy-momentum tensor at second-order provides a correction to the one presented in \cite{Baumann:2007zm} as Eq. 5, where they did not correctly account for the coupling between the scalar-perturbation and the anisotropic stress. For instance, we see that \eqref{second_order_spatial_energy_tensor} agrees with Eq. 4.17 from Ref. \cite{Malik:2008im}.}
\begin{equation}
\label{second_order_spatial_energy_tensor}
    T^{(2)i}_{j} = P^{(2)}\delta^{i}_{j} + 2\left(\bar{\rho} + \bar{P}\right)v^{i(1)}v^{(1)}_{j} + a^{-2}\left[\pi^{(2)i}_{j} + \left(4\firstpsi\delta^{ik}\right)\pi^{(1)}_{jk}\right]\ .
\end{equation}
The evolution of second-order tensor perturbations can be found by applying the projection tensor $P^{li}_{jm}$ \cite{Carbone:2004iv} to the second-order Einstein's equation Eq.~\eqref{eq: second-order EFE}. We extract the transverse-traceless parts of Eqs.~\eqref{second_order_spatial_einstein_tensor}-\eqref{second_order_spatial_energy_tensor} and obtain 
\begin{align}
\label{second_order_spatial_einstein_tensor_projected}
    P^{li}_{jm}G^{(2)m}_{l} &= a^{-2}\left[\frac{1}{4}h^{i(2)''}_{j} + \frac{1}{2}\frac{a'}{a}h^{i(2)'}_{j} - \frac{1}{4}\nabla^{2}h^{i(2)}_{j} + \partial^{i}\firstphi\partial_{j}\firstphi + 2\firstphi\partial^{i}\partial_{j}\firstphi \right.\nonumber\\
    & \left. - 2\firstpsi\partial^{i}\partial_{j}\firstphi - \partial_{j}\firstphi\partial^{i}\firstpsi - \partial^{i}\firstphi\partial_{j}\firstpsi + 3\partial^{i}\firstpsi\partial_{j}\firstpsi + 4\firstpsi\partial^{i}\partial_{j}\firstpsi\right]\ , \numberthis
\end{align}
\begin{equation}
\label{second_order_spatial_energy_tensor_projected}
    P^{li}_{jm}T^{(2)m}_{l} = \left(\bar{\rho} + \bar{P}\right)v^{i(1)}v^{(1)}_{j} + a^{-2}\left(4\firstpsi\delta^{ik}\pi^{(1)}_{jk}\right)\ . 
\end{equation}
Replacing the first-order expressions of the four-velocity and of the anisotropic stress, Eq.~\eqref{second_order_spatial_energy_tensor_projected} can be recast as
\begin{align*}
\label{second_order_spatial_energy_tensor_projected_2}
    P^{li}_{jm}T^{(2)m}_{l} = & \frac{4}{3\mathcal{H}^{2}\left(1 + w\right)}\left[\partial^{i}\left(\psi^{(1)'} + \mathcal{H}\firstphi\right)\partial_{j}\left(\psi^{(1)'} + \mathcal{H}\firstphi\right)\right] \\
    & - 4\firstpsi\delta^{ik}\left[\left(\partial_{j}\partial_{k} - \frac{1}{3}\nabla^{2}\delta_{jk}\right)\left(\firstphi - \firstpsi\right)\right]\ , \numberthis
\end{align*}
where $w\equiv \bar{P}/\bar{\rho}$ is the equation-of-state parameter. Then, Eq. (\ref{eq: second-order EFE}) becomes
\begin{equation}
\label{einstein_eq_projection}
    h^{i(2)''}_{j} + 2\mathcal{H}h^{i(2)'}_{j} - \nabla^{2}h^{i(2)}_{j} = -4P^{li}_{jm}S_{l}^{m}\ , 
\end{equation}
where the source term is given by 
\begin{align*}
\label{sourceterm_poisson_gr}
    P^{li}_{jm}S_{l}^{m} & = \partial^{i}\firstphi\partial_{j}\firstphi + 2\firstphi\partial^{i}\partial_{j}\firstphi - 2\firstpsi\partial^{i}\partial_{j}\firstphi - \partial_{j}\firstphi\partial^{i}\firstpsi - \partial^{i}\firstphi\partial_{j}\firstpsi \\
    & + 3\partial^{i}\firstpsi\partial_{j}\firstpsi + 4\firstpsi\partial^{i}\partial_{j}\firstpsi \\
    & - \frac{4}{3\mathcal{H}^{2}\left(1 + w\right)}\left[\partial^{i}\left(\psi^{(1)'} + \mathcal{H}\firstphi\right)\partial_{j}\left(\psi^{(1)'} + \mathcal{H}\firstphi\right)\right] \\
    & + 4\firstpsi\delta^{ik}\left[\left(\partial_{j}\partial_{k} - \frac{1}{3}\nabla^{2}\delta_{jk}\right)\left(\firstphi - \firstpsi\right)\right]\ . \numberthis
\end{align*}
We know that the scalar perturbations satisfy the following constraint equation 
\begin{equation}
\label{constraint_equation_anisotropic_stress}
	\left(\partial^{i}\partial_{j} - \frac{1}{3}\nabla^{2}\delta^{i}_{j}\right)\left(\firstphi - \firstpsi\right) = \kappa^{2}\pi^{(1)i}_{j}\ .
\end{equation}
Thus, in the absence of anisotropic stress, we have $\firstpsi =\firstphi$ and the source term (\ref{sourceterm_poisson_gr}) reduces to \cite{Ananda:2006af}
\begin{align*}
\label{sourceterm_poisson_gr_simplified}
    P^{li}_{jm}S_{l}^{m} & = 4\firstphi\partial^{i}\partial_{j}\firstphi + \frac{2(1+3w)}{3(1+w)}\partial^{i}\firstphi\partial_{j}\firstphi  \\
    & - \frac{4}{3\mathcal{H}^{2}\left(1 + w\right)}\left(\partial^{i}\dfirstphi\partial_{j}\dfirstphi + \mathcal{H}\partial^{i}\dfirstphi\partial_{j}\firstphi + \mathcal{H}\partial^{i}\firstphi\partial_{j}\dfirstphi\right)\ . \numberthis
\end{align*}
Eq.~\eqref{einstein_eq_projection} can be solved in Fourier space, where the source term becomes a convolution of the first-order scalar perturbations with different wave-vectors $\mathbf{k}$. The tensor mode can be re-written in Fourier space as 
\begin{equation}
h^{(2)}_{ij}(\mathbf{x},\tau)=\sum_{\lambda=+,\times}\dfrac{1}{(2\pi)^{3/2}}\int d^3\mathbf{k}\; e^{i\mathbf{k}\cdot\mathbf{x}} h^{(2)}_{\lambda}(\mathbf{k},\tau)e^\lambda_{ij}(\mathbf{k})\;,
    \label{eq: fourier transform of GW}
\end{equation}
where $\lambda=+,\times$ denotes the two GW polarizations and $e^\lambda_{ij}(\mathbf{k})$ are the polarization tensors. In Fourier space, the GW equation (\ref{einstein_eq_projection}) results \footnote{We have added the subscript GR in Eq.~\eqref{wave_equation_second_order_tensor_mode} to distinguish it from the contribution that will be computed in the next section in $f(R)$ gravity.}
\begin{equation}
\label{wave_equation_second_order_tensor_mode}
    h^{(2)''}_\lambda(\mathbf{k},\tau) + 2\mathcal{H}h^{(2)'}_\lambda(\mathbf{k},\tau) + k^{2}h^{(2)}_\lambda(\mathbf{k},\tau) = S_{\lambda, \rm GR}(\mathbf{k},\tau)\,, 
\end{equation}
where the source term $S_{\lambda,\rm GR}(\mathbf{k},\tau)$ depends on the evolution of the first-order scalar mode $\firstphi(\mathbf{k},\tau)$. The potential $\firstphi(\mathbf{k},\tau)$ obeys the following equation of motion (E.o.M)
\begin{equation}
\label{simplified_first_order_evolution_equation}
    \phi^{(1)''} + 3\mathcal{H}\left(1 + c_{s}^{2}\right)\phi^{(1)'} + \left(2\mathcal{H}' + \left(1 + 3c_{s}^{2}\right)\mathcal{H}^{2} + c_{s}^{2}k^{2}\right)\firstphi = 0\ ,
\end{equation}
where $c_{s}^{2} \equiv P^{(1)}/\rho^{(1)}$ is the adiabatic speed of sound.  
In the case of a perfect fluid and in the absence of entropy perturbations $c_{s}^{2} = w$ and the E.o.M reduces to the following
\begin{equation}
\label{simplified_first_order_evolution_equation fourier}
    \phi^{(1)''}(\mathbf{k},\tau) + 3\mathcal{H}\left(1 + w\right)\phi^{(1)'}(\mathbf{k},\tau) + w k^{2}\firstphi(\mathbf{k},\tau) = 0\ .
\end{equation}
From (\ref{simplified_first_order_evolution_equation fourier}), notice that on super-horizon scales (i.e. $k\ll \mathcal{H}$), independently on $w$, $\phi^{(1)}(\mathbf{k},\tau)$ sets to a constant. Hence, $\phi_\mathbf{k}(\tau)$ can be split into the initial condition $\zeta_\mathbf{k}$ set by inflation and a transfer function $T_\phi(k\tau)$, encoding the time-dependence acquired by the modes once they re-enter the horizon in the subsequent phases of the evolution of the universe.
In full generality, we can write
\begin{equation}
\label{split_scalar_mode}
    \firstphi (\mathbf{k},\tau) = \frac{3(1+w)}{5+3w}T_{\phi}(k\tau)\zeta_{\mathbf{k}}\ .
\end{equation}
The primordial curvature fluctuations $\zeta_{\mathbf{k}}$ are related to the power spectrum $\mathcal{P}_\zeta(k)$ as 
\begin{equation}
\label{fluctuation_power_spectrum}
    \langle \zeta_{\mathbf{k}}\zeta_{\mathbf{q}} \rangle = \delta^{(3)}(\mathbf{k} + \mathbf{q})\mathcal{P}_{\zeta}(k)\ , 
\end{equation}
where the power spectrum can be written in terms of the dimensionless power spectrum $\Delta^2_\zeta(k)$ as
\begin{equation}
    \mathcal{P}_\zeta(k)=\dfrac{2\pi^2}{k^3}\Delta^2_\zeta(k) \ .
\end{equation}
Considering radiation domination, i.e. $w=1/3$, and solving Eq.  (\ref{simplified_first_order_evolution_equation fourier}), the transfer function of the scalar potential reads
\begin{equation}
\label{approximate_solution_phi_radiation}
    T_\phi (k\tau) = \frac{9}{(k\tau)^{2}}\left[\frac{\sqrt{3}}{k\tau}\sin\left(\frac{k\tau}{\sqrt{3}}\right) - \cos\left(\frac{k\tau}{\sqrt{3}}\right)\right]\ .
\end{equation} 
Using the splitting~\eqref{split_scalar_mode} for the scalar mode, $S_{\lambda, \rm GR}(\mathbf{k},\tau)$ can be written as \cite{Domenech:2021ztg}
\begin{align*}
\label{fourier_sourceterm}
    S_{\lambda, \rm GR}(\mathbf{k},\tau) & = -4 e^{ij}_{\lambda}(\mathbf{k})S_{ij}(\mathbf{k}, \tau)\\
    & =  4\int \frac{d^3\mathbf{q}}{(2\pi)^{3/2}}Q_{\lambda}(\mathbf{k}, \mathbf{q})f(|\mathbf{k} - \mathbf{q}|,q,\tau)\zeta_{\mathbf{q}}\zeta_{\mathbf{k} - \mathbf{q}}\ ,  \numberthis
\end{align*}
where we have introduced the projection factor, $Q_{\lambda}(\mathbf{k}, \mathbf{q}) \equiv \mathbf{e}^{ij}_{\lambda }(\mathbf{k})\mathbf{q}_{i}\mathbf{q}_{j}$, which encodes the polarization tensors, and the source function
\begin{align*}
\label{source_function_expanded}
    f(|\mathbf{k} - \mathbf{q}|,q,\tau) & \equiv \frac{3(1+w)}{(5+3w)^{2}}\left\{2(5+3w)T_{\phi}(q\tau)T_{\phi}(|\mathbf{k} - \mathbf{q}|\tau)  + \frac{4}{\mathcal{H}^{2}}T'_{\phi}(q\tau)T'_{\phi}(|\mathbf{k} - \mathbf{q}|\tau)\right. \\
    & \left. + \frac{4}{\mathcal{H}}\left[T_{\phi}(q\tau)T'_{\phi}(|\mathbf{k} - \mathbf{q}|\tau) + T_{\phi}(q\tau)T'_{\phi}(|\mathbf{k} - \mathbf{q}|\tau)\right]\right\}, \numberthis
\end{align*}
which includes the time evolution in the transfer functions and the dependence on the equation of state $w$.

\section{\label{sec:second_order_mg}SIGW in beyond-GR theories}

So far, we have worked under the assumption that the Universe is described by small inhomogeneous perturbations on a flat background, with the two scalar modes, $\firstphi$ and $\firstpsi$, satisfying the constraint equation ~\eqref{constraint_equation_anisotropic_stress}. Going to Fourier space, the source of the anisotropic stress contribution is given by \cite{Mangilli:2008bw}
\begin{equation}
\label{anisotropic_stress}
	k^{2}\left(\firstpsi - \firstphi\right) = -4\kappa^{2}a^{2}\left[\rho_{\gamma}\Theta_{2} + \rho_{\nu}\mathcal{N}_{2}\right], 
\end{equation}
where $\Theta_{2}$ and $\mathcal{N}_{2}$ refer to the quadrupole moment of free-streaming photons and neutrinos, respectively. As discussed above, in GR we can safely assume that both contributions on the right side of Eq.~\eqref{anisotropic_stress} are negligibly small, such that $\firstpsi \simeq \firstphi$.
However, this is not necessarily true in modified theories of gravity, in which a \textit{gravitational slip}, i.e. a difference between the two scalar potentials, can be sourced by extra terms in the Einstein-Hilbert action due to purely geometrical effects. In this case, non-standard GW propagation effects have been studied in Ref. \cite{DeFelice:2011hq,Saltas:2014dha,Sawicki:2016klv}. In this case, the anisotropy constraint equation can be generically written as \cite{Saltas:2014dha}
\begin{equation}
\label{constraint_eq_modified}
	\firstpsi - \firstphi = \sigma(a)\Pi + \pi_{m}\ , 
\end{equation} 
where $\Pi$ depends on the background and linear order quantities, $\sigma(a)$ is a background function and $\pi_{m}$ is the anisotropic contribution due to matter fields, which is non-vanishing for imperfect fluids. 
Thus, we now seek to explore the physics of induced GWs within a general theory of gravity. Specifically, we briefly review $f(R)$ gravity \cite{DeFelice:2010aj} in the metric formalism and our analytical setup. We provide the background equations, the evolution equations of linear cosmological perturbations and the equation of motion of SIGW in a generic $f(R)$ model.
To determine the changes that beyond-GR terms introduce in the SIGW power spectral density we consider a specific model, i.e. $f(R)=R+\alpha R^2$.

\subsection{\label{subsec:f_R} \texorpdfstring{$f(R)$ gravity}{f(R) gravity}}

In $f(R)$ gravity, the Lagrangian density $f$ is a general function of the Ricci scalar $R$. Hence, the 4-dimensional action reads (see the reviews \cite{DeFelice:2010aj,Tsujikawa:2010zza,Sotiriou:2008rp} and references therein)
\begin{equation}
\label{action_modifed_gravity}
    S = \frac{1}{2\kappa^2}\int d^{4}x\sqrt{-g}f(R) + \int d^{4}x\mathcal{L}_{M}, 
\end{equation}
where $g$ is the determinant of the metric $g_{\mu\nu}$, $\mathcal{L}_M$ is the matter Lagrangian and in general, it depends on the metric $g_{\mu\nu}$ and on the matter fields. 
The field equations can be derived by varying the action~\eqref{action_modifed_gravity} with respect to $g_{\mu\nu}$:
\begin{equation}
\label{field_equation_1}
     F(R)R_{\mu\nu} - \frac{1}{2}f(R)g_{\mu\nu} - \nabla_{\mu}\nabla_{\nu}F(R) + g_{\mu\nu}\Box F(R)=\kappa^{2}T_{\mu\nu} \ , 
\end{equation}
where $F(R) \equiv \partial f / \partial R$.  
The trace of \eqref{field_equation_1} gives
\begin{equation}
    \label{trace of field equation_1}
    3\Box F(R)+F(R)R-2f(R)=\kappa^2T
\end{equation}
where $T=g^{\mu\nu}T_{\mu\nu}$. We know that $f(R)=R$ and $F(R)=1$ in the case of GR without a cosmological constant. Therefore, it is clear from the above equation that $R$ is given directly by $T$ in GR as $R=-\kappa^2T$. However, for a general form of $f(R)$, $\Box F(R)\not= 0$, thus Eq. \eqref{trace of field equation_1} determines the dynamics of a propagating scalar field $\varphi\equiv F(R)$, the so-called \textit{scalaron} \cite{Starobinsky:1980te}.\\
Eq. ~\eqref{field_equation_1} can also be recast in terms of the standard Einstein tensor, $G_{\mu\nu}$, by replacing 
\begin{align*}
\label{modified_einstein_tensor}
    F(R)G_{\mu\nu} & = F(R) R_{\mu\nu} - \frac{1}{2}g_{\mu\nu}F(R)R\ .
\end{align*}
such that
\begin{equation}
\label{field_equation_2}
    {F(R)}G_{\mu\nu} = g_{\mu\nu}\frac{(f(R)-RF(R))}{2} + \nabla_{\mu}\nabla_{\nu}F(R) - g_{\mu\nu}\Box F(R) + \kappa^{2}T_{\mu\nu}\,.
\end{equation}

Hereafter, we consider a specific model, i.e., $f(R) = R + \alpha R^{2}$, which was originally proposed by Starobinsky to explain cosmological inflation \cite{Starobinsky:1980te}. Such a model is consistent with the temperature anisotropies and E-mode polarization observed in the CMB~\cite{Planck:2018jri}. As for the GR case, we consider perturbations of $f(R)$ gravity around a spatially flat FLRW space and expand all quantities at linear order in the coupling $\alpha$, such that 
\begin{align*}
    f(R) & = \bar{f} + f^{(1)} + f^{(2)} = R^{(0)} + R^{(1)} + R^{(2)} + \alpha 
    \left(R^{(0)} + R^{(1)} +  R^{(2)}\right)^{2}\ , \\
    F(R) & = \bar{F} + F^{(1)} + F^{(2)} = 1 + 2\alpha \left(R^{(0)} + R^{(1)} + R^{(2)}\right)\ . 
\end{align*} 
where the overline denotes the background quantities and the apex indicates the order of the perturbations.
This is because, since we are dealing with a general theory of gravity deep in the radiation-dominated era, we expect that the corrections introduced with respect to GR are subdominant.
Hence, we can make a perturbative expansion of the scalar modes around their GR solution in the following way \cite{article}
\begin{equation}
\label{modified_phi_def}
    \firstphi = \firstphigr + \alpha\,\delta\firstphi\,,
\end{equation}
\begin{equation}
\label{modified_psi_def}
    \firstpsi = \firstpsi_{\text{GR}} + \alpha\,\delta\firstpsi\,,
\end{equation}
where $\delta\firstpsi$ and $\delta\firstphi$ are the beyond-GR corrections at first-order. In the absence of anisotropic stress in GR, we have $\firstphigr = \firstpsi_{\text{GR}}$, such that $\firstpsi = \firstphigr + \alpha\delta\firstpsi$. Similarly, pressure and energy-density can be expanded as 
\begin{equation}
\label{modified_rho_def}
    \rho^{(1)} = \rho^{(1)}_{\text{GR}} + \alpha\,\delta\rho^{(1)},
\end{equation}
\begin{equation}
\label{modified_p_def}
    P^{(1)} = P^{(1)} _{\text{GR}} + \alpha\,\delta P^{(1)}\ . 
\end{equation}
By applying these decompositions to the field equations iteratively by the parameter $\alpha$, we can determine quantitatively the beyond-GR corrections of a given $f(R)$ model and a reference model (i.e. GR). Clearly, we recover GR for $\alpha\rightarrow0$. For the remainder, we do not take into consideration the first-order $\alpha$ corrections to the background quantities $a$ and $\mathcal{H}$, due to negligible contribution (refer to Appendix \ref{sec:appendixB} for details). 

\subsection{\label{subsec:first_order_modified}Modified first-order field equations}

To study beyond-GR effects in the power spectrum of SIGWs we need to understand the behavior of the scalar perturbations $\firstphi$ and $\firstpsi$. Hence, we perturbed the field equations (\ref{field_equation_2}) at linear order. From the time-time component, we get the following energy constraint, 
\begin{align*}
\label{generic_modified_00_component_first_order}
    3\mathcal{H}(\mathcal{H}\firstphi & + \dfirstpsi)  - \nabla^{2}\firstpsi = \\
    & =  -\frac{1}{2\bar{F}}\left[a^{2}\kappa^{2}\rho^{(1)} + 3\mathcal{H}'F^{(1)} - 3\mathcal{H}F^{(1)'} + \nabla^{2}F^{(1)} + 3(2\mathcal{H}\firstphi + \dfirstpsi)\bar{F}'\right]\ ,\numberthis
\end{align*}
which for $f(R)=R+\alpha R^2$ reduces to
\begin{align*}
\label{modified_00_component_first_order}
    & 6\mathcal{H}^2\firstphi + 6\mathcal{H}\dfirstpsi  - 2\nabla^{2}\firstpsi + \kappa^{2}a^{2}\rho^{(1)} = \\
    & = 2a^{-2}\alpha\left[ \left(108\mathcal{H}^4 + 36(\mathcal{H}')^{2} - 72\mathcal{H}''\mathcal{H}\right)\firstphi + \left(24\mathcal{H}^{2} + 18\mathcal{H}' + 2\nabla^{2}\right)\nabla^{2}\firstphi -4\nabla^{2}\nabla^{2}\firstpsi\right. \\
    & \left. - 12\mathcal{H}^{2}\nabla^{2}\firstpsi - 36\mathcal{H}'\mathcal{H}\dfirstphi  + \left(108\mathcal{H}^{3} - 36\mathcal{H}'\mathcal{H} - 18\mathcal{H}''\right)\dfirstpsi  + 30\mathcal{H}\nabla^{2}\dfirstpsi \right. \\
    & \left. - 18\mathcal{H}^{2}\ddfirstphi   + \left(18\mathcal{H}' - 18\mathcal{H}^{2}\right)\ddfirstpsi  + 6\nabla^{2}\ddfirstpsi  - 18\mathcal{H}\dddfirstpsi \right], \numberthis
\end{align*}
The momentum constraint is found from the time-space component of the field equations:
\begin{equation}
\label{generic_modified_momentum_constraint}
     \partial_{i}\dfirstpsi + \mathcal{H}\partial_{i}\firstphi = \frac{1}{2\bar{F}}\left[-a^{2}\kappa^{2}(\bar{\rho}+\bar{P})v_{i}^{(1)} - \mathcal{H}\partial_{i}F^{(1)} + \partial_{i}F^{(1)'} - \partial_{i}\firstphi\bar{F}'\right].
\end{equation}
From the above equation, we find an expression for the velocity components for the case $f(R)=R+\alpha R^2$
\begin{align*}
\label{modified_four_velocity}
     v_{i}^{(1)} & = -\frac{2}{a^{2}\kappa^{2}\left(\bar{\rho} + \bar{P}\right)}\left(\mathcal{H}\partial_{i}\firstphi + \partial_{i}\dfirstpsi\right)  \\
    & - \frac{\alpha}{a^{4}\kappa^{2}\left(\bar{\rho} + \bar{P}\right)}\partial_{i}\left([36\mathcal{H}'' - 72\mathcal{H}^{3}]\firstphi + [36\mathcal{H}' - 12\mathcal{H}^{2}]\dfirstphi + [60\mathcal{H}' - 84\mathcal{H}^{2}]\dfirstpsi\right. \\
     & \left. + 12\mathcal{H}\ddfirstphi  + 12\ddddfirstpsi  + 12\mathcal{H}\left(2\nabla^{2}\firstpsi - \nabla^{2}\firstphi\right) + 4\left(\nabla^{2}\dfirstphi - 2\nabla^{2}\dfirstpsi\right)\right). \numberthis
\end{align*}
Finally, we compute the trace and trace-free components of the spatial field equation.\\
The trace reads 
\begin{align*}
\label{generic_modified_ij_trace_component_first_order}
    \ddfirstpsi + 2\mathcal{H}\dfirstpsi + \mathcal{H}\dfirstphi + (2\mathcal{H}' + \mathcal{H}^{2})\firstphi & = \frac{1}{2\bar{F}}\left[a^{2}\kappa^{2}P^{(1)} + \frac{2}{3}\bar{F}\nabla^{2}(\firstpsi - \firstphi) \right. \\
    & \left. - \frac{2}{3}\nabla^{2}F^{(1)}  - (\mathcal{H}' + 2\mathcal{H}^{2})F^{(1)} + \mathcal{H}F^{(1)'} - F^{(1)''} \right. \\
    & \left. - 2\firstphi\bar{F}'' - (2\mathcal{H}\firstphi + \dfirstphi + 2\dfirstpsi)\bar{F}'\right]\ , \numberthis
\end{align*}
which in $f(R)=R+\alpha R^2$ can be recast as
\begin{align*}
\label{modified_ij_trace_component_first_order}
   & 2\mathcal{H}\dfirstphi  + \left[4\mathcal{H}' + 2\mathcal{H}^{2}\right]\firstphi + \frac{2}{3}\nabla^{2}\firstphi + 4\mathcal{H}\dfirstpsi  + 2\ddfirstpsi  - \frac{2}{3}\nabla^{2}\firstpsi - \kappa^{2}a^{2}P^{(1)}= \\
    & = 2a^{-2}\alpha\left[\left(144\mathcal{H}'\mathcal{H}^{2} + 24\mathcal{H}''\mathcal{H} - 12(\mathcal{H}')^{2} - 24\mathcal{H}''' - 36\mathcal{H}^{4}\right)\firstphi \right. \\
    & \left. + \left(10\mathcal{H}' + 4\mathcal{H}^{2} + \frac{4}{3}\nabla^{2}\right)\nabla^{2}\firstphi + \left(4\mathcal{H}^{2}-8\mathcal{H}'-\frac{8}{3}\nabla^{2}\right)\nabla^{2}\firstpsi \right. \\
    & \left. + \left(12\mathcal{H}'\mathcal{H} - 36\mathcal{H}'' + 36\mathcal{H}^{3}\right)\dfirstphi  + 10\mathcal{H}\nabla^{2}\dfirstphi  \right. \\
    & \left. + \left(84\mathcal{H}'\mathcal{H} - 30\mathcal{H}''\right)\dfirstpsi   + \left(6\mathcal{H}^{2} - 24\mathcal{H}'\right)\ddfirstphi  - 2\nabla^{2}\ddfirstphi  \right. \\
    & \left. + \left(42\mathcal{H}^{2} - 30\mathcal{H}'\right)\ddfirstpsi  + 8\nabla^{2}\ddfirstpsi  - 6\mathcal{H}\dddfirstphi  - 6\ddddfirstpsi \right]\ . \numberthis
\end{align*}
The trace-free component is given by:
\begin{equation}
\label{generic_modified_tracefree}
    \left(\partial^{i}\partial_{j}-\frac{1}{3}\nabla^{2}\delta^{i}_{j}\right)\left(\firstpsi - \firstphi - \frac{F^{(1)}}{\bar{F}}\right) = \frac{1}{\bar{F}}\kappa^{2}\pi^{(1)i}_{j}\ .
\end{equation}
The gravitational slip is encoded in Eq.~\eqref{generic_modified_tracefree}, which can be expanded at the linear order in $\alpha$ as
\begin{align*}
\label{modified_anisotropic_stress}
    \kappa^{2}\pi^{(1)i}_{j} &= -\left(1 + 12\alpha a^{-2}\left[\mathcal{H}'+\mathcal{H}^{2}\right]\right)\left(\partial^{i}\partial_{j} - \frac{1}{3}\nabla^{2}\delta^{i}_{j}\right)\left(\firstphi-\firstpsi\right) \\
    & - 2\alpha a^{-2}\left(\partial^{i}\partial_{j} - \frac{1}{3}\nabla^{2}\delta^{i}_{j}\right)\left( - 6\ddfirstpsi  - 6\mathcal{H}\dfirstphi  - 18\mathcal{H}\dfirstpsi  \right. \\
     & \left. - 12\left[\mathcal{H}'+\mathcal{H}^{2}\right]\firstphi - 2\nabla^{2}\firstphi + 4\nabla^{2}\firstpsi\right). \numberthis
\end{align*}
As previously discussed, the anisotropic stress is sourced by an imperfect fluid depending on the matter properties which is not affected by the beyond-GR modifications and we can still neglect its (small) contribution. Hence, we can set $\pi^{(1)i}_{j}=0$ and this allows us to find a relation between $\delta\firstpsi$ and $\delta\firstphi$. The Eqs.~\eqref{modified_00_component_first_order}, ~\eqref{modified_ij_trace_component_first_order} and ~\eqref{modified_anisotropic_stress} can be expanded using Eqs.~\eqref{modified_phi_def} -~\eqref{modified_p_def}. In $f(R)$ gravity thus, the two scalar potentials differ, and from ~\eqref{modified_anisotropic_stress} one finds
\begin{equation}
\label{delta_psi_definition}
    \delta\firstpsi = \delta\firstphi + 2a^{-2}\left(-6\ddfirstphigr - 24\mathcal{H}\dfirstphigr - 12\left[\mathcal{H}' + \mathcal{H}^{2}\right]\firstphigr + 2\nabla^{2}\firstphigr\right).
\end{equation}
Using the definition $P^{(1)} = c_{s}^{2}\rho^{(1)}$, we can relate Eq.~\eqref{modified_00_component_first_order} to Eq.~\eqref{modified_ij_trace_component_first_order} in radiation epoch, where $c_{s}^{2} = 1/3$. After some manipulations, we find the following evolution equation for $\delta\firstphi$:
\begin{align*}
\label{modified_evolution_equation_scalar_perturbation}
    & \delta\ddfirstphi + 4\mathcal{H}\delta\dfirstphi + 2\left[\mathcal{H}' + \mathcal{H}^{2}\right]\delta\firstphi - \frac{1}{3}\nabla^{2}\delta\firstphi = \\
    & = \frac{2}{3} a^{-2} \left[\left(36\mathcal{H}''\mathcal{H} -72\mathcal{H}^{4}  + 6\mathcal{H}^{2}\nabla^{2} + \nabla^{4}\right)\firstphigr + \left(36\mathcal{H}'' - 72\mathcal{H}^{3} - 12\mathcal{H}\nabla^{2}\right)\dfirstphigr \right. \\
    & \left. + \left(72\mathcal{H}' - 18\mathcal{H}^{2} - 6\nabla^{2}\right)\ddfirstphigr + 36\mathcal{H}\dddfirstphigr + 9\ddddfirstphigr\right], \numberthis
\end{align*}
By substituting $a(\tau)=a_\ast(\tau/\tau_\ast)$ and $\mathcal{H} = \tau^{-1}$ into Eq.~\eqref{modified_evolution_equation_scalar_perturbation} and going to Fourier space, we find
\begin{align*}
\label{first_order_evolution_equation_radiation_mg}
    \delta\ddfirstphi + 4\tau^{-1}\delta\dfirstphi + \frac{1}{3}k^{2}\delta\firstphi & = \left(a_{*}\frac{\tau}{\tau_{*}}\right)^{-2}\left[\left( \frac{2}{3}k^{4}-4\tau^{-2}k^{2}\right)\firstphigr + 8\tau^{-1}k^{2}\dfirstphigr \right. \\
    & \left. + ( 4k^{2}-60\tau^{-2})\ddfirstphigr + 24\tau^{-1}\dddfirstphigr + 6\ddddfirstphigr\right]\ . \numberthis
\end{align*}
\noindent 
where $\tau_\ast$ and $a_\ast$ refer to some reference time.
Eq. \eqref{first_order_evolution_equation_radiation_mg} is the evolution equation for $\delta\phi^{(1)}$, which we will need to solve in order to obtain the source term of the E.o.M. of SIGW.
Before proceeding, let us stress an important point. Notice that it is not necessary to use the first-order evolution of the scalaron degree of freedom \footnote{Notice that Eq. \eqref{trace of field equation_1} is nothing but a wave equation for $F(R)$. The field equation \eqref{trace of field equation_1} for the additional propagating scalar degree of freedom at linear order gives \begin{equation}
F^{(1)''}+2\mathcal{H}F^{(1)'}-(\nabla^2-a^2m^2_s)F^{(1)}=\dfrac{a^2\kappa^2}{3}T+(4\mathcal{H}\phi^{(1)}\bar{F}'+2\phi^{(1)}\bar{F}''+\phi^{(1)'}\bar{F}'+3\psi^{(1)'}\bar{F}')
\end{equation} where the scalaron mass squared reads $m^2_s\equiv\dfrac{1}{3}\left(\dfrac{\bar{F}}{\bar{F}_{,R}}-R\right)$, where $\bar{F}_{,R}\equiv d\bar{F}/dR$ (see, e.g., \cite{DeFelice:2010aj,Zhou:2024doz}). Notice that in the above equation, no particular form of $f(R)$ has been specified, and no expansion around a spatially flat FLRW space has been performed.} to close the system of equations for the fluctuations. This is similar to what happens in Ref. \cite{Acquaviva:2002ud}. Nevertheless, we have reported the evolution equation at linear order in footnote (3) for completeness.

\subsection{\label{subsec:second_order_modified}Modified second-order field equations}

In order to find the source term of the ``scalar-induced'' gravitational waves in $f(R)$ gravity, we study the trace-free and transverse component of the second-order spatial field equation (\ref{field_equation_2}). We find the following generic expression
\begin{align*}
\label{generic_trace_free_transverse_spatial_second_order}
     \ddtensorh + \left(2\mathcal{H}+ \frac{\bar{F}'}{\bar{F}}\right)\dtensorh - \nabla^{2}\tensorh = - \frac{4}{\bar{F}}P^{li}_{jm}S^{m}_{l,\text{MG}}, \numberthis
\end{align*}
where $S^{i}_{j,\text{MG}}$ is 
\begin{align*}
\label{generic_source_term_modified}
    P^{li}_{jm}S^{m}_{l,\text{MG}} & = \left(\partial^{i}\firstphi\partial_{j}\firstphi + 2\firstphi\partial^{i}\partial_{j}\firstphi - 2\firstpsi\partial^{i}\partial_{j}\firstphi - \partial_{j}\firstphi\partial^{i}\firstpsi - \partial^{i}\firstphi\partial_{j}\firstpsi \right. \\
    & \left. + 3\partial^{i}\firstpsi\partial_{j}\firstpsi + 4\firstpsi\partial^{i}\partial_{j}\firstpsi\right)\bar{F} - \partial^{i}\partial_{j}F^{(2)} + \partial^{i}\partial_{j}(\firstpsi - \firstphi)F^{(1)} \\
    & - 2\firstpsi\partial^{i}\partial_{j}F^{(1)} - \partial^{i}\firstpsi\partial_{j}F^{(1)} - \partial_{j}\firstpsi\partial^{i}F^{(1)} - a^{2}\kappa^{2}T^{(2)i}_{j}\ ,\numberthis
\end{align*}
and the second-order energy-momentum tensor can be found by substituting Eq.~\eqref{modified_four_velocity} and Eq.~\eqref{modified_anisotropic_stress} into Eq.~\eqref{second_order_spatial_energy_tensor_projected}. By taking into account the modified energy density background quantity, Eq.~\eqref{modified_background_00}, the E.o.M. of SIGW becomes
\begin{align*}
\label{second_order_tensor_equation_mg}
     \frac{1}{4}\left(1 +  12a^{-2}\alpha\left[\mathcal{H}' + \mathcal{H}^{2}\right]\right) & \left(\ddtensorh + 2\mathcal{H}\dtensorh - \nabla^{2}\tensorh\right) \\
     & + 3\alpha a^{-2}\left[\mathcal{H}'' - 2\mathcal{H}^{3}\right]\left(h^{(2)i}_{j}\right)' = S_{j}^{i} + \alpha\delta S_{j}^{i}\ , \numberthis
\end{align*}
where $\alpha\delta S_{j}^{i}$ is the modified source term. The full source term is presented in Appendix \ref{sec:appendixC}. As done in section \ref{subsec:evolution_equation}, we write Eq.~\eqref{second_order_tensor_equation_mg} in Fourier space and we evaluate it in the radiation-domination era. Thus, we get
\begin{equation}
\label{second_order_tensor_modified_radiation}
    h^{(2)''}_\lambda(\mathbf{k},\tau) + 2\tau^{-1}h^{(2)'}_\lambda(\mathbf{k},\tau) + k^{2}h^{(2)}_\lambda(\mathbf{k},\tau) = S_{\lambda}(\mathbf{k},\tau) = S_{\lambda,\rm GR}(\mathbf{k},\tau) + \alpha\delta S_{\lambda}(\mathbf{k},\tau)\ ,
\end{equation}
where $S_{\lambda, \rm GR}(\mathbf{k},\tau)$ is given by Eq.~\eqref{fourier_sourceterm} and, similarly, we write $\delta S_{\lambda}(\mathbf{k},\tau)$ as
\begin{align*}
\label{fourier_source_term_modified_radiation}
    \delta S_{\lambda}(\mathbf{k},\tau) & = -4\mathbf{e}^{ij}_{\lambda}(\mathbf{k})\delta S_{ij}(\mathbf{k},\tau) \\
     &= 4\int \frac{d^3\mathbf{\mathbf{q}}}{(2\pi)^{3/2}}Q_{\lambda}(\mathbf{k}, \mathbf{q})\delta f(|\mathbf{k} - \mathbf{q}|,q,\tau)\zeta_{\mathbf{q}}\zeta_{\mathbf{k} - \mathbf{q}}\ .  \numberthis
\end{align*}
The beyond-GR correction to the source function reads
\begin{align*}
\label{modifed_source_function_radiation}
    \delta f(|\mathbf{k} - \mathbf{q}|,q,\tau) & = \frac{4}{9}\Bigl\{3T_{\delta\phi}(q\tau)T_{\phi}(|\mathbf{k}-\mathbf{q}|\tau) + 3T_{\phi}(q\tau)T_{\delta\phi}(|\mathbf{k}-\mathbf{q}|\tau) \\
    &  + \tau T'_{\delta\phi}(q\tau)T_{\phi}(|\mathbf{k}-\mathbf{q}|\tau) + \tau T_{\delta\phi}(q\tau)T'_{\phi}(|\mathbf{k}-\mathbf{q}|\tau)  \\
    &  + \tau T'_{\phi}(q\tau)T_{\delta\phi}(|\mathbf{k}-\mathbf{q}|\tau) + \tau T_{\phi}(q\tau)T'_{\delta\phi}(|\mathbf{k}-\mathbf{q}|\tau) \\
    &  + \tau^{2} T'_{\phi}(q\tau)T'_{\delta\phi}(|\mathbf{k}-\mathbf{q}|\tau) + \tau^{2} T'_{\delta\phi}(q\tau)T'_{\phi}(|\mathbf{k}-\mathbf{q}|\tau) \\
    &  - \left(\frac{\tau_{*}}{a_{*}}\right)^{2}\tau^{-2}\left[- 96T'_{\phi}(q\tau)T'_{\phi}(|\mathbf{k}-\mathbf{q}|\tau) \right. \\
    &  \left.  + 30\left(T_{\phi}(\mathbf{q}\tau)T''_{\phi}(|\mathbf{k}-\mathbf{q}|\tau) + T''_{\phi}(\mathbf{q}\tau)T_{\phi}(|\mathbf{k}-\mathbf{q}|\tau)\right) \right.  \\
    &  \left.  + 6\tau\left(T_{\phi}(\mathbf{q}\tau)T'''_{\phi}(|\mathbf{k}-\mathbf{q}|\tau) + T'''_{\phi}(\mathbf{q}\tau)T_{\phi}(|\mathbf{k}-\mathbf{q}|\tau)\right)  \right. \\
    &  \left. + 18\tau\left(T'_{\phi}(q\tau)T''_{\phi}(|\mathbf{k}-\mathbf{q}|\tau) + T''_{\phi}(\mathbf{q}\tau)T'_{\phi}(|\mathbf{k}-\mathbf{q}|\tau)\right) \right. \\
    &  \left.  + 6\tau^{2}\left(T'_{\phi}(q\tau)T'''_{\phi}(|\mathbf{k}-\mathbf{q}|\tau) + T'''_{\phi}(\mathbf{q}\tau)T'_{\phi}(|\mathbf{k}-\mathbf{q}|\tau)\right) \right.  \\
    &  \left. + 2\left(\mathbf{q}^{2} + |k-\mathbf{q}|^{2}\right)T_{\phi}(\mathbf{q}\tau)T_{\phi}(|\mathbf{k}-\mathbf{q}|\tau)\right. \\
    &  \left.  + 2\tau^{2}\left(\mathbf{q}^{2} + |k-\mathbf{q}|^{2}\right)T'_{\phi}(q\tau)T'_{\phi}(|\mathbf{k}-\mathbf{q}|\tau) \right. \\
    &  \left.  + 2\tau\mathbf{q}^{2}\left(T'_{\phi}(q\tau)T_{\phi}(|\mathbf{k}-\mathbf{q}|\tau) - T_{\phi}(\mathbf{q}\tau)T'_{\phi}(|\mathbf{k}-\mathbf{q}|\tau)\right) \right.  \\
    &  \left.  + 2\tau|k-\mathbf{q}|^{2}\left(T_{\phi}(\mathbf{q}\tau)T'_{\phi}(|\mathbf{k}-\mathbf{q}|\tau) - T'_{\phi}(q\tau)T_{\phi}(|\mathbf{k}-\mathbf{q}|\tau)\right) \right] \Bigr\} \ .  \numberthis
\end{align*}
To get the expression above, we used Eq.~\eqref{second_order_source_term_modified_expanded} in Fourier space and we have defined the split of the scalar fluctuation as 
\begin{equation}
\label{modified_split_scalar_mode}
    \firstphi_{\mathbf{k}} (\tau) = \firstphi_{\mathbf{k},\text{GR}}+ \alpha \delta\firstphi_{\mathbf{k}} = \frac{3(1+w)}{5+3w}\left(T_{\phi}(k\tau) + \alpha T_{\delta\phi}(k\tau) \right)\zeta_{\mathbf{k}}.
\end{equation}
where $T_{\delta\Phi}(k\tau)$ comes from solving Eq.~\eqref{first_order_evolution_equation_radiation_mg}.
\noindent
From Eq.~\eqref{first_order_evolution_equation_radiation_mg}, we note that we cannot use the analytical solution Eq.~\eqref{approximate_solution_phi_radiation} for the first-order scalar perturbation in the GR case, since in pure radiation we would have $R^{(0)}= R^{(1)}=0$. Therefore, we should solve the E.o.M. for $\delta\phi^{(1)}$ using the full GR solution including the sub-dominant matter content in RD. 
However, we do not include subdominant matter perturbations in the computation of the GR contribution. This simplification is made to facilitate further calculations, as the cold dark matter correction to the scalar potential in GR contributes only subdominantly to the source term and has a negligible effect on the kernel and resulting energy spectrum. To quantify the impact of this approximation, we compared the numerical kernel including cold dark matter corrections to the one computed without them. The mean percentage difference was found to be as low as $9 \times 10^{-10}\%$, with a maximum deviation of only $3.2 \times 10^{-7}\%$ across the full range of $x\equiv k\tau$. These negligible differences confirm the validity of neglecting cold dark matter perturbations in the GR kernel within our analysis. A detailed treatment of matter perturbations during the matter-dominated era, including their inclusion in GR, can be found in Ref.~\cite{Kohri:2018awv,Domenech:2021ztg}.

\subsection{\label{subsec:matter_correction_solution}Matter correction in standard GR and modified scalar potential}

We consider the sub-dominant matter content during the radiation era. To obtain the matter correction to the scalar potential in the radiation-dominated era, we write the first-order energy constraint as 
\begin{align*}
\label{first_order_einstein_eq_time_GR}
    -\nabla^{2}\firstpsi + 3\mathcal{H}\left(\dfirstpsi + \mathcal{H}\firstphi\right) & = -\frac{\kappa^{2}a^{2}}{2}\rho^{(1)}, \\
    & = -\frac{\kappa^{2}a^{2}}{2}\left(\bar{\rho}_{m}\delta_{m} + \bar{\rho_{r}}\delta_{r}\right), \numberthis
\end{align*}
where in the second equality we have introduced the density contrast $\delta = \rho^{(1)}/\bar{\rho}$ and the subscripts \textit{m} and \textit{r} denote the matter and radiation contribution, respectively. Similarly, considering the two-component fluid, the momentum constraint reads
\begin{align*}
\label{first_order_einstein_eq_time_space_GR}
    \mathcal{H}\partial_{i}\firstphi + \partial_{i}\dfirstpsi & = -\frac{\kappa^{2}a^{2}}{2}\left(\bar{\rho} + \bar{P}\right)v^{(1)}_{i}, \\
    & = -\frac{\kappa^{2}a^{2}}{2}\left(\bar{\rho} + \bar{P}\right)(\partial_{i}v + v_{i}^{V}), \\
    \mathcal{H}\firstphi + \dfirstpsi & = -\frac{\kappa^{2}a^{2}}{2}\left\{\bar{\rho}_{m}\left(1 + w_{m}\right)\frac{\theta_{m}}{\nabla^{2}} + \bar{\rho}_{r}\left(1 + w_{r}\right)\frac{\theta_{r}}{\nabla^{2}}\right\}, \numberthis
\end{align*}
where in the second step we have decomposed the velocity into its scalar and vector parts as $v_{i} = \partial_{i}v + v_{i}^{V}$ with $\partial^iv_{i}^{V}=0$. In the last step, we have extracted the longitudinal part and introduced the velocity divergence as $\theta = \nabla^{2}v$. 
Using Eq.~\eqref{first_order_einstein_eq_time_space_GR},~\eqref{first_order_einstein_eq_time_GR} reads 
\begin{align*}
\label{correction_phi_equation}
    \nabla^{2}\firstpsi & = \frac{\kappa^{2}a^{2}}{2}\left(\bar{\rho}_{m}\delta_{m} + \bar{\rho_{r}}\delta_{r} - \bar{\rho}_{m}\left(1 + w_{m}\right)\frac{\theta_{m}}{\nabla^{2}} - \bar{\rho}_{r}\left(1 + w_{r}\right)\frac{\theta_{r}}{\nabla^{2}}\right), \\
    & = \frac{3}{2}\mathcal{H}^{2}\left(\frac{\tau}{\tau_{\text{eq}}}\left[\delta_{m} - \left(1 + w_{m}\right)\frac{\theta_{m}}{\nabla^{2}}\right] + \delta_{r} - \left(1 + w_{r}\right)\frac{\theta_{r}}{\nabla^{2}}\right) \numberthis   \;,
\end{align*}
where in the second equality we have used the Friedmann equation, $\mathcal{H}^{2} = \frac{a^{2}\kappa^{2}}{3}\bar{\rho}_{r}$, and the relation $\frac{\bar{\rho}_{m}}{\bar{\rho}_{r}} = \frac{\tau}{\tau_{\text{eq}}}$. Non-relativistic matter is subdominant in the radiation era, thus we use the solution for the scalar potential Eq.~\eqref{approximate_solution_phi_radiation} serves as the external source for $\delta_{m}$ and $\theta_{m}$. We solve for $\delta_{m}$ and $\theta_{m}$ using the first-order perturbed energy-momentum conservation equations \cite{Dodelson:2003ft}
\begin{equation}
\label{density_contrast}
    \delta' + 3\mathcal{H}(c_{s}^{2} - w)\delta = - (1+w)(\theta - 3\dfirstpsi)\,,
\end{equation}
\begin{equation}
\label{velocity}
    \theta' + \left[\mathcal{H}(1-3w) + \frac{w'}{1+w}\right]\theta = -\nabla^{2}\left(\frac{c_{s}^{2}}{1+w}\delta + \firstphi\right).
\end{equation}
Going to Fourier space, we then solve Eq.~\eqref{correction_phi_equation} for $\phi$ using $\firstphi=\firstpsi$. We find that the matter correction reads
\begin{align*}
\label{approximate_solution_phi_matter_correction}
    \firstphi_{\text{GR,m}} & = \frac{9}{4}\frac{1}{ k \tau_{\text{eq}} x^3} \left(-3 x^2 \left(2 \text{Ci}\left(\frac{x}{\sqrt{3}}\right)-2 \log (x)+\log (3)\right)+(6 \gamma -3) x^2 \right. \\
    & \left. +6 \sqrt{3} x \sin \left(\frac{x}{\sqrt{3}}\right)+18 \cos \left(\frac{x}{\sqrt{3}}\right)-18\right), \numberthis
\end{align*}
where $\gamma \approx 0.577$ is the Euler–Mascheroni constant and $\tau_{\text{eq}} \approx 10^{16}$ is the time of radiation-matter equality. The resulting modified scalar potential solution is found by solving Eq.~\eqref{first_order_evolution_equation_radiation_mg} using~\eqref{approximate_solution_phi_matter_correction} on the RHS, 
\begin{align*}
\label{postGR_scalar_solution}
    \delta\firstphi & = \frac{27}{2}\frac{1}{\tau_{\text{eq}}}\frac{1}{x^5}k^3 \left(\frac{\tau_{*}}{a_{*}}\right)^2\left(\left(x^2-6\right) \left(-2 \text{Ci}\left(\frac{x}{\sqrt{3}}\right)+2 \log (x)+2 \gamma -1-\log (3)\right) \right. \\
    & \left. + 2 \sqrt{3} x \sin \left(\frac{x}{\sqrt{3}}\right)-6 \cos \left(\frac{x}{\sqrt{3}}\right)\right). \numberthis
\end{align*}
This allows us to have an analytical expression for the modified source function Eq.~\eqref{modifed_source_function_radiation}.

\section{\label{sec:power_spectrum_modified}Power spectrum of induced GWs}

We now compute the power spectrum of the scalar-induced GWs for both the GR and beyond-GR parts. The solution $h^{(2)}_\lambda(\mathbf{k},\tau)$ of \eqref{second_order_tensor_modified_radiation} is found using Green's method as
\begin{equation}
\label{solution_h}
    h^{(2)}_\lambda(\mathbf{k},\tau) = \frac{1}{a(\tau)}\int_{\tau_{i}}^{\tau} d\Tilde{\tau} G_{\mathbf{k}}(\tau,\Tilde{\tau})a(\Tilde{\tau})S_\lambda(\mathbf{k},\Tilde{\tau}) , 
\end{equation} 
where $G_{\mathbf{k}}(\tau,\Tilde{\tau})$ is the Green's function of the homogeneous equation and in the radiation era it is given by 
\begin{equation}
\label{greens_function}
    G_\mathbf{k}(\tau, \Tilde{\tau}) = \frac{1}{k}\sin(k\tau-k\Tilde{\tau})\ . 
\end{equation}
The power-spectrum of tensor modes is defined as \cite{Matarrese:1993zf}
\begin{equation}
\label{power_spectrum_definition}
 	\langle h_{\lambda}(\mathbf{k},\tau)h_{\lambda'}(\mathbf{k'},\tau) \rangle = \frac{2\pi^{2}}{k^{3}}\delta_{\lambda\lambda'}\delta^{(3)}(\mathbf{k}+\mathbf{k'}) \Delta^2_{h,\lambda}(k,\tau)\ .
\end{equation}
Using Eqs.~\eqref{fourier_sourceterm} and \eqref{fourier_source_term_modified_radiation}, we can write the two-point correlation function for the tensor modes in the following way
\begin{align*}
\label{tensor_two_point_correlation_function}
    \langle h_{\lambda_1}(\mathbf{k_1},\tau) h_{\lambda_2}(\mathbf{k_2},\tau) \rangle & =  \frac{1}{a^{2}(\tau)} \int_{\tau_{i}}^{\tau} d\Tilde{\tau_{1}}\int_{\tau_{i}}^{\tau} d\Tilde{\tau_{2}} G_{\mathbf{k_{1}}}(\tau,\Tilde{\tau_{1}})G_{\mathbf{k_{2}}}(\tau,\Tilde{\tau_{2}})a(\Tilde{\tau_{1}})a(\Tilde{\tau_{2}})\\
    & \times \langle \Bigl(S_{\lambda_{1},\text{GR}}(\mathbf{k_{1}},\Tilde{\tau_{1}}) + \alpha\delta S_{\lambda_{1}}(\mathbf{k_{1}},\Tilde{\tau_{1}})\Bigr) \Bigl(S_{\lambda_{2},\text{GR}}(\mathbf{k_{2}},\Tilde{\tau_{2}}) + \alpha\delta S_{\lambda_{2}}(\mathbf{k_{2}},\Tilde{\tau_{2}})\Bigr) \rangle\ . \numberthis   
\end{align*}
Expanding the correlator up to the first order in $\alpha$, we have
\begin{align*}
\label{two_point_correlation_function_source}
    \langle S_{\lambda_{1},\text{GR}}(\mathbf{k_{1}},\tau_{1})S_{\lambda_{2},\text{GR}}({\mathbf{k_{2}}},\tau_{2}) \rangle  & =  16 \int \frac{d^3\mathbf{q_{1}}}{(2\pi)^{3/2}}\int \frac{d^3\mathbf{q_{2}}}{(2\pi)^{3/2}} Q_{{\lambda_{1}}}(\mathbf{k_{1}}, \mathbf{q_{1}})Q_{{\lambda_{2}}}(\mathbf{k_{2}}, \mathbf{q_{2}}) \\
    & \times f(|\mathbf{k_{1}} - \mathbf{q_{1}}|,q_{1},\tau)f(|\mathbf{k_{2}} - \mathbf{q_{2}}|,q_{2},\tau)\langle \zeta_{\mathbf{q_{1}}}\zeta_{\mathbf{k_{1}} - \mathbf{q_{1}}}\zeta_{\mathbf{q_{2}}}\zeta_{\mathbf{k_{2}} - \mathbf{q_{2}}} \rangle\ , \numberthis   
\end{align*}
and 
\begin{align*}
\label{two_point_correlation_function_source_modifed}
    \alpha \langle \delta S_{\lambda_{1}}({\mathbf{k_{1}}},\tau_{1})S_{\lambda_{2},\text{GR}}({\mathbf{k_{2}}},\tau_{2}) \rangle  & =  \,\alpha \,16 \int \frac{d^3\mathbf{q_{1}}}{(2\pi)^{3/2}}\int \frac{d^3\mathbf{q_{2}}}{(2\pi)^{3/2}} Q_{{\lambda_{1}}}(\mathbf{k_{1}}, \mathbf{q_{1}})Q_{{\lambda_{2}}}(\mathbf{k_{2}}, \mathbf{q_{2}}) \\
    & \times \delta f(|\mathbf{k_{1}} - \mathbf{q_{1}}|,q_{1},\tau)f(|\mathbf{k_{2}} - \mathbf{q_{2}}|,q_{2},\tau)\langle \zeta_{\mathbf{q_{1}}}\zeta_{\mathbf{k_{1}} - \mathbf{q_{1}}}\zeta_{\mathbf{q_{2}}}\zeta_{\mathbf{k_{2}} - \mathbf{q_{2}}} \rangle\ . \numberthis   
\end{align*}
The contribution from $\alpha \langle  S_{\lambda_{1},\text{GR}}({\mathbf{k_{1}}},\tau_{1})\delta S_{\lambda_{2}}({\mathbf{k_{2}}},\tau_{2}) \rangle$ is similar to Eq. \eqref{two_point_correlation_function_source_modifed}.
The 4-point correlation function can be decomposed into a disconnected contribution, i.e., the product of two 2-point functions, and a connected one, which will vanish if the primordial curvature perturbations are Gaussian-distributed \cite{Cai:2018dig,Perna:2024ehx}. In this case, the 4-point function  becomes \cite{Domenech:2021ztg} 
\begin{align*}
\label{scalar_four_point_function}
   \langle \zeta_{\mathbf{q_1}}\zeta_{\mathbf{k_1}-\mathbf{q_1}}\zeta_{\mathbf{q_2}}\zeta_{\mathbf{k_2}-\mathbf{q_2}}\rangle & = \frac{2\pi^{2}}{{q_1}^{3}}\Delta^2_{\zeta}({q_1})\frac{2\pi^{2}}{|\mathbf{k_1}-\mathbf{q_1}|^{3}}\Delta^2_{\zeta}(|\mathbf{k_1}-\mathbf{q_1}|)\times\\
   & \times \delta^{(3)}(\mathbf{q_1}+\mathbf{q_2})\delta^{(3)}(\mathbf{k_1}+\mathbf{k_2}-\mathbf{q_1}-\mathbf{q_2}) + (\mathbf{q_2}\leftrightarrow\mathbf{k_2}-\mathbf{q_2})\ . \numberthis
\end{align*}
It is convenient to work in terms of the dimensionless variables $u\equiv\frac{|\mathbf{k}-\mathbf{q}|}{k}$ and $v\equiv\frac{{q}}{k}$ as done e.g. in Ref. \cite{Kohri:2018awv}. 
Using Eq.~\eqref{power_spectrum_definition} and \eqref{two_point_correlation_function_source}-\eqref{scalar_four_point_function}, from Eq. \eqref{tensor_two_point_correlation_function} we obtain
\begin{align*}
\label{power_spectrum_spherical_coordinates}
    \Delta^2_{h}(k,\tau) = & 8 \int^{\infty}_{0} dv \int^{1+v}_{|1-v|} du \left(\frac{4v^2 - \left(1-u^2+v^2\right)^2}{4uv}\right)^{2} \\
    & \times \left(I^{2}_{\text{GR}}(v,u,x) + 2\alpha\delta I(v,u,x)I_{\text{GR}}(v,u,x)\right)\Delta^2_{\zeta}(ku)\Delta^2_{\zeta}(kv)\ , \numberthis
\end{align*}
where we defined $x \equiv k \tau$ and we already summed over the two GW polarizations, so ${\Delta^2_{h}(k,\tau)}=\sum_{\lambda=+,\times}{\Delta^2_{h,\lambda}(k,\tau)}$.
Moreover, we have defined the kernel in GR and beyond-GR as 
\begin{equation}
\label{kernel_radiation}
   I_{\text{GR}}(|\mathbf{k} - \mathbf{q}|,q,\tau) \equiv \int_{\tau_{i}}^{\tau} d\tilde{\tau}\; \frac{a(\tilde{\tau}) }{a(\tau)}G_{\mathbf{k}}(\tau, \tilde{\tau}) f(|\mathbf{k} - \mathbf{q}|,q,\tau),
\end{equation}
and
\begin{equation}
\label{modified_kernel_radiation}
    \delta I(|\mathbf{k} - \mathbf{q}|,q,\tau)  \equiv \int_{\tau_{i}}^{\tau} d\tilde{\tau}\; \frac{a(\tilde{\tau}) }{a(\tau)}G_{\mathbf{k}}(\tau, \tilde{\tau}) \delta f(|\mathbf{k} - \mathbf{q}|,q,\tau),
\end{equation}
respectively. 
Since we are interested in modes deeply inside the horizon, $x \gg 1$, we take the late-time limit, i.e. $\tau \rightarrow \infty$. The GR kernel \eqref{kernel_radiation} can be evaluated analytically and it reads \cite{Kohri:2018awv}:
\begin{align*}
\label{kernel_radiation_analytic}
    I_{RD}(v,u,x \rightarrow \infty) = & \frac{3(u^{2}+v^{2}-3)}{4u^{3}v^{3}x}\left(\sin{x}\left(-4uv + (u^{2} + v^{2} - 3)\log\left|\frac{3-(u+v)^{2}}{3-(u-v)^{2}}\right|\right) \right. \\
    & \left. -\pi(u^{2}+v^{2}-3)\Theta(u+v-\sqrt{3})\cos{x}\right), \numberthis
\end{align*} 
where $\Theta$ is the Heaviside theta function. 
However, to compute the spectral density of GWs we need the time-averaged power spectrum, so one needs to compute the oscillation average of the square of Eq.~\eqref{kernel_radiation_analytic} \cite{Kohri:2018awv}:
\begin{align*}
\label{oscillation_avg_kernel}
    \overline{I_{RD}^{2}(v,u,x \rightarrow \infty)} = & \frac{1}{2}\left(\frac{3(u^{2}+v^{2}-3)}{4u^{3}v^{3}x}\right)^{2}\left(\left(-4uv + (u^{2} + v^{2} - 3)\log\left|\frac{3-(u+v)^{2}}{3-(u-v)^{2}}\right|\right)^{2} \right. \\
    & \left. -\pi^{2}(u^{2}+v^{2}-3)^{2}\Theta(u+v-\sqrt{3})\right). \numberthis
\end{align*} 
The beyond-GR kernel at first-order in $\alpha$, i.e. Eq.~\eqref{modified_kernel_radiation}, can be found using Eq.~\eqref{modifed_source_function_radiation}. 
Notice that there are two parts in Eq.~\eqref{modifed_source_function_radiation}; the first four lines contain the contribution arising from the coupling between the beyond-GR and GR scalar fluctuations, while the remaining lines are purely GR contributions. As we are only considering the first-order leading terms, we make the following choice. We use Eq.~\eqref{approximate_solution_phi_radiation} for $\firstphi$ and Eq.~\eqref{postGR_scalar_solution} for $\delta\firstphi$ when considering the beyond-GR and GR coupling. On the other hand, for the rest of Eq.~\eqref{modifed_source_function_radiation}, we already pointed out how the coupling of pure-radiation solutions for $\firstphi$ gives a vanishing contribution; instead, we decide to couple the matter correction solution, i.e. Eq.~\eqref{approximate_solution_phi_matter_correction}, and the radiation solution \eqref{approximate_solution_phi_radiation}. We check that the terms arising from the coupling of the matter correction solutions together are subdominant compared to the coupling between pure-radiation and matter solutions.

\begin{figure}[t!]
    \centering  
    \includegraphics[width=0.7\textwidth]{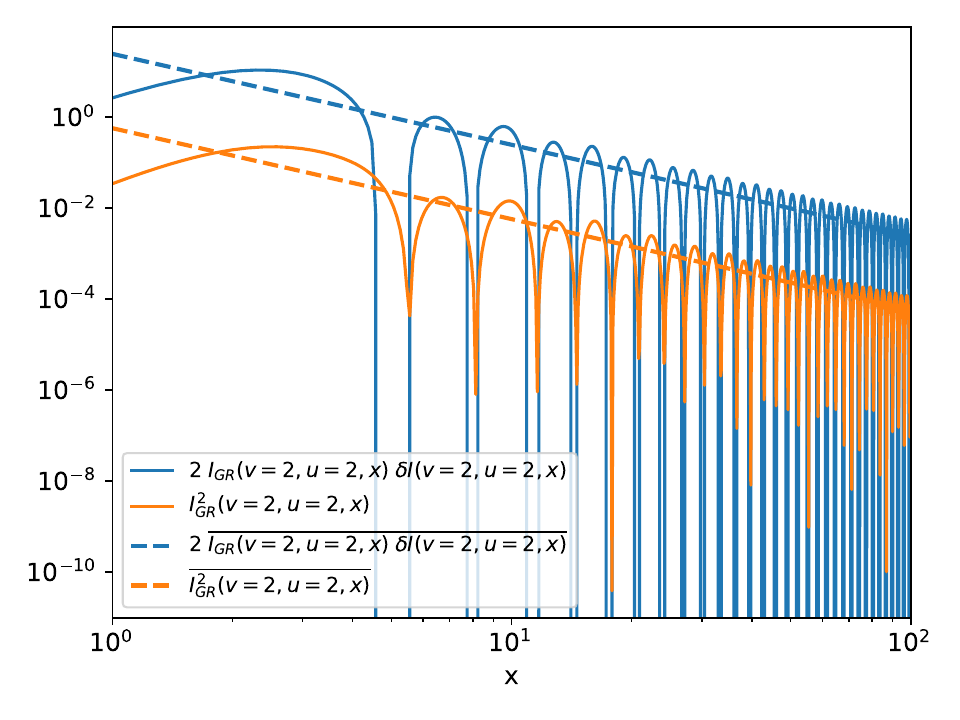}  
    \caption{GR and beyond-GR 
    correction to the kernel function as a function of $x\equiv k\tau$ and fixing a pair of $v$ and $u$ values. The 
    GR contribution (thick orange line), i.e. $I^{2}_{\text{GR}}(v=2,u=2,x)$, is computed using Eq.~\eqref{kernel_radiation}, while the oscillation average (dashed orange line) is obtained from Eq.~\eqref{oscillation_avg_kernel}. The numerical integration of the beyond-GR component (thick blue line) is computed using $2I_{\text{GR}}(v=2,u=2,x)\delta I(v=2,u=2,x)$, i.e. Eqs.~\eqref{kernel_radiation} and ~\eqref{modified_kernel_radiation}, and its oscillation average is drawn as dashed blue line. Since the MG contribution is negative, we plot it here with a negative sign, in order to facilitate the comparison with the GR result.}  
    \label{fig:numerical_kernel}  
\end{figure}

In this way, given the expression for $\delta f$, we perform a numerical integration of the kernel \eqref{modified_kernel_radiation} by fixing a pair of $u$ and $v$ values and we compare it to the GR kernel \eqref{kernel_radiation} in
Fig.~\ref{fig:numerical_kernel}. 

\section{Induced GW spectrum in $f(R)$ gravity}
\label{sec:spectral_energy_density}
\setcounter{figure}{0}

Using the expression of the overall SIGW power spectrum taking into consideration standard GR and the first-order $\alpha$ correction in $f(R)$ gravity, i.e. Eq. \eqref{power_spectrum_spherical_coordinates}, we can now study the observable GW spectral energy density. The averaged energy density of GWs is given by \cite{Espinosa:2018eve}
\begin{equation}
\label{GW_energy_density}
    \rho_{\rm GW}(\tau) = \frac{1}{16a^{2}(\tau)\kappa^2}\overline{\langle(\nabla h_{ij})^{2}\rangle}\,, 
\end{equation}
where the overbar denotes the oscillation average. Hence, the fractional energy density of GWs per logarithmic wavenumber interval reads \cite{Adshead:2021hnm}
\begin{equation}
\label{SIGW_fractional_energy_density}
    \Omega_{\rm GW}(k,\tau) \equiv \frac{\rho_{\rm GW}(k,\tau)}{\rho_{c}(\tau)} = \frac{1}{48}\left(\frac{k}{a(\tau)H(\tau)}\right)^{2}\sum_{\lambda=+,\times}\overline{{\Delta^2_{h,\lambda}(k,\tau)}}\ ,
\end{equation}
where $\rho_{c}(\tau)$ is the critical energy density. Given the expression Eq.~\eqref{power_spectrum_spherical_coordinates} for the SIGW power spectrum, we can split the spectral energy density as $\Omega_{\rm GW}(k,\tau) = \Omega_{\rm GW,\text{GR}}(k,\tau) + 2 \alpha\,\delta\Omega_{\rm GW}(k,\tau)$, where $\Omega_{\text{GR,GW}}$ and $\delta \Omega_{\text{GW}}$ are the GR and beyond-GR contribution, respectively. It is worth noting that, in modified gravity theories such as \(f(R)\), the effective Planck mass \(M_{\rm eff}^2(t)\) can in general be time-dependent (see, e.g., \cite{Dalang:2019rke}). In the present analysis, the parameter \(\alpha\) is treated as a constant. Incorporating a time-dependent \(\alpha\), and thus a fully dynamical \(M_{\rm eff}^2(t)\), is a possible extension of this work and may provide further insights into the evolution of the gravitational wave energy density in \(f(R)\) gravity frameworks.
 
Assuming that the GW emission occurs after the reheating phase, we can relate the spectral density of GWs during the radiation era to the GW spectral density at present $\Omega_{GW,0}(k)$, which is relevant for observations, as follows \cite{Watanabe:2006qe}:
\begin{equation}
    \Omega_{\rm GW,0}(k)h^{2} = \Omega_{r,0}(k)h^{2}\left(\frac{g_{*s}(T_{hc})}{g_{*s0}}\right)^{-1/3}\left(\frac{g_{*}(T_{hc})}{g_{*0}}\right)\Omega_{\rm GW,hc}(k)\ ,
\end{equation}
where $h = H_{0}/100\,\text{km}\,\text{s}^{-1}\,\text{Mpc}^{-1}$ is the reduced Hubble constant and $g_{*}$ and $g_{*s}$ are the effective number of relativistic degrees of freedom. In the above expression, the subscript "hc" means that the quantity is evaluated at the ’horizon crossing’ epoch, i.e. the emission of GWs, while the subscript "0" denotes the present-day values. 
Recall that in the SM $g_\ast(T_{hc})=g_{\ast,S}(T_{hc})=106.75$ for $T_{hc}\gtrsim1\;\rm{TeV}$ \cite{Watanabe:2006qe}. 
$\Omega_{r,0}$ corresponds to the present value of the energy density fraction of radiation, which is constrained to be $\Omega_{r,0}h^{2} \sim 4.2 \times 10^{-5}$ \cite{Planck:2018vyg}. 
We compute the spectral energy density in the frequency ranges of current- and future-generation detectors focusing on: (i) the Pulsar Timing Array (PTA) with $f_{\rm PTA} \sim 10^{-8} \,\text{Hz}$, (ii) the Laser Interferometer Space Antenna (LISA) with $f_{\rm LISA} \sim 10^{-3} \,\text{Hz}$ and (iii) the LIGO-Virgo-KAGRA (LVK) interferometer network with $f_{\rm LVK} \sim 10^{2} \,\text{Hz}$.

For the purpose of this work, we consider a dimensionless seed curvature power spectrum with a Gaussian peak in logarithmic $k$-space~\cite{Pi:2020otn},
\begin{equation}
\label{scalar_power_spectrum}
    \Delta^2_{\zeta}(k) = \frac{\mathcal{A}_{\zeta}}{\sqrt{2\pi\sigma^{2}}}\exp\left(-\frac{\ln^{2}({k/k_{*}})}{2\sigma^{2}}\right), 
\end{equation}
where the amplitude $\mathcal{A}_{\zeta}$ satisfies the normalization $\int^\infty_0 \Delta^2_{\zeta}(k) d \ln k ={\cal A}_\zeta$, $\sigma$ is the dimensionless width and $k_{*}$ is the peak wave number. For the numerical integration of the energy density, we take $\sigma=0.1$ and $\mathcal{A}_\zeta = 2.1\times 10^{-2}$, which is larger than the constraint at the CMB scale, i.e. $k_{CMB}=0.05\;\rm{Mpc}^{-1}$, $A_\zeta(k_{CMB})\sim 2.1 \times 10^{-9}$ (TT, TE, EE+low-E+lensing \cite{Planck:2018jri}, \cite{Planck:2018vyg}). This is because we do not have such tight constraints of the scalar power spectrum on smaller scales, therefore, in principle, the CMB bounds are not applicable at the scales of interest without making large extrapolations.
We fix $ k_{*}$ depending on the reference scale of the detector we are considering.

\subsection{\texorpdfstring{Numerical calculation of SIGW spectrum in $f(R)$ gravity}{Numerical calculation of SIGW spectrum in f(R) gravity}}

\begin{figure}[t!]
\centering
\includegraphics[width=0.7\textwidth]{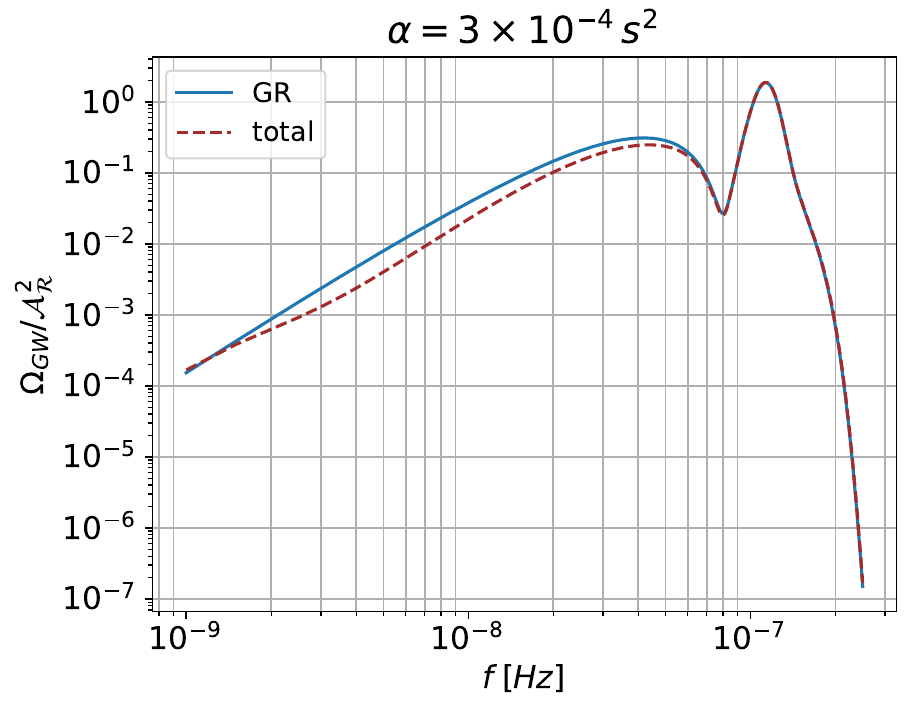}
    \caption{SIGW spectrum for PTA scales, assuming $k_{*} = 6.25 \times 10^{7}\, \text{Mpc}^{-1}$ and for an example value of $\alpha$. We use a log-normal seed with $\sigma = 0.1$, normalized with respect to the amplitude $\mathcal{A}_{\zeta}^{2}$. 
    The thick blue line indicates the GR SIGW spectrum, while the dashed red line indicates the total SIGW spectrum including the first-order correction in $\alpha$, as explained in the main text.} 
    \label{fig:PTA_spectrum}
\end{figure}

\begin{figure}[htbp]
    \centering
        \includegraphics[width=0.7\textwidth]{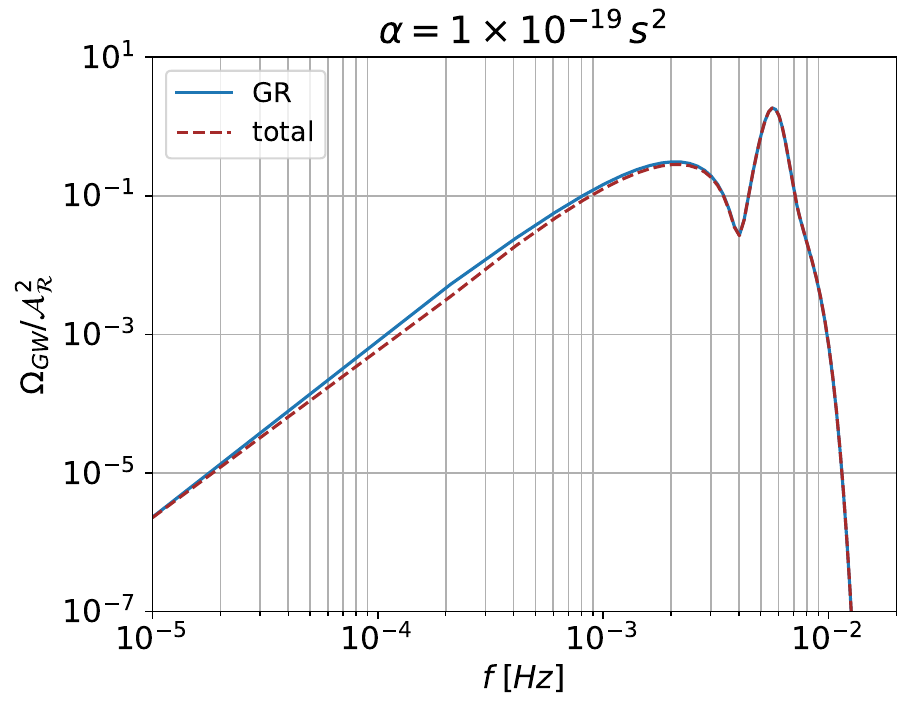}

    \caption{Same as Fig.~\ref{fig:PTA_spectrum} but for the LISA detector by fixing $k_\ast=3.23\times 10^{12}\;\rm Mpc^{-1}$ and considering an example value of $\alpha$.}
    \label{fig:LISA_spectrum}
\end{figure}

\begin{figure}[htbp]
    \centering
        \includegraphics[width=0.7\textwidth]{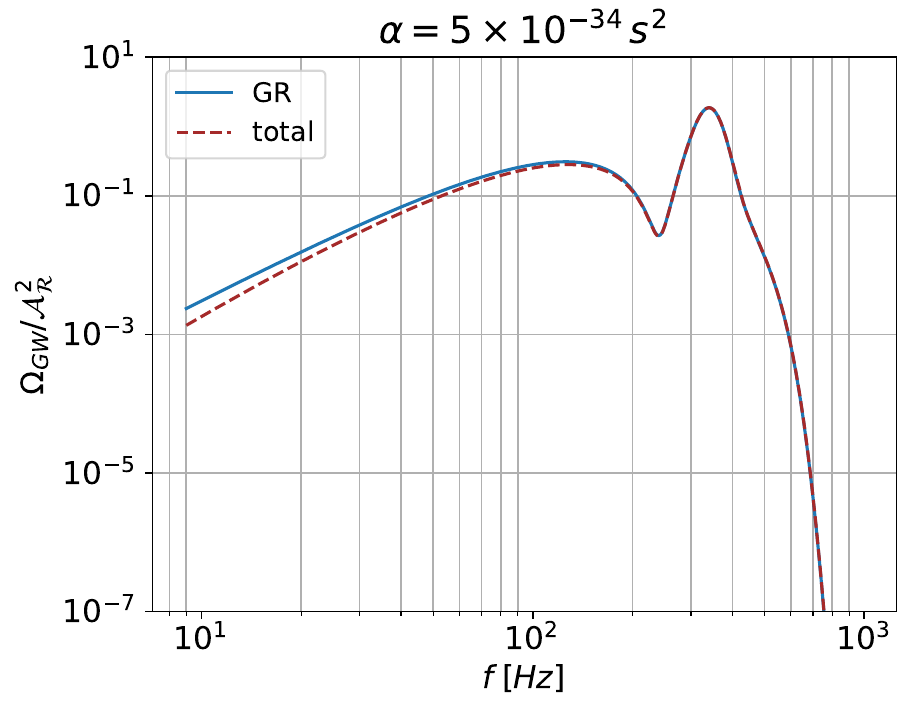}
    \caption{Same as Fig.~\ref{fig:PTA_spectrum} but for the LVK network by fixing $k_{*} = 1.96 \times 10^{17} \,\text{Mpc}^{-1}$ and considering an example value of $\alpha$.}  
    \label{fig:LVK_spectrum}
\end{figure}

Here, we estimate the impact of the beyond-GR effect on the SIGW spectrum. 
%Given our framework, we determine the SIGWs spectral energy density as $\Omega_{\text{GW}} =\Omega_{\text{GR,GW}}+2\alpha\delta \Omega_{\text{GW}}$, where $\Omega_{\text{GR,GW}}$ and $\delta \Omega_{\text{GW}}$ are the GR and beyond-GR contribution, respectively. 
The ratio between the beyond-GR and GR contribution to the spectral density scales as $\propto\,\alpha \left(\tau_{*}/a_{*}\right)^{2} k^{3}/\tau_{\text{eq}}$. In order to estimate such an effect, we have to constrain $\alpha$ and $\left(\tau_{*}/a_{*}\right)^{2}$. We constrain the latter by using the background Friedmann equation deep in the radiation era. We find  that $(\tau_{*}/a_{*})^{2} \simeq 10^{40}\,  \text{s}^{2} $. Given the value for $\tau_{eq} \simeq 10^{16}\,  \text{s}$, and considering  $k = 2\pi f$ for each detector, we get the following pre-factor values: (i) for PTA is $\propto \alpha \times 10^{2} \, \text{s}^{-2}$, (ii) for LISA is $\propto \alpha \times 10^{17} \, \text{s}^{-2}$ and (iii) for LVK is $\propto \alpha \times 10^{32} \, \text{s}^{-2}$. Given this pre-factor, we proceed to numerically integrate the SIGW energy spectrum in GR and beyond-GR, $\Omega_{\text{GR,GW}}(k,\tau)$ and $\delta\Omega_{\rm GW}(k,\tau)$. 
For the GR solution, we retain the solution in pure radiation, while the MG contribution is primarily determined by the subdominant matter contribution present during the radiation-dominated era. We present the SIGW spectrum for PTA, LISA and LVK frequencies in Figures \ref{fig:PTA_spectrum}, \ref{fig:LISA_spectrum} and \ref{fig:LVK_spectrum}, respectively, assuming a value of $\alpha$ which guarantees to be under perturbative regime and to have a positive GW energy density, since we want that the beyond-GR contribution is smaller than the GR contribution.  

From Figures \ref{fig:PTA_spectrum}, \ref{fig:LISA_spectrum} and \ref{fig:LVK_spectrum}, we observe that the low-frequency tail of the SIGW spectrum gets suppressed by the MG contribution, which thus acts as a damping term. 
Notice that, for stability reasons, $\alpha$ needs to be positive since it is related to the mass squared of the scalar degree of freedom introduced in $f(R)$ gravity \cite{DeFelice:2010aj}.
We argue that observational data from PTA, LISA and LVK could constraint the value of $\alpha$ on scales on which we have limited information. 
Moreover, even if we are taking $\alpha$ to be large at the time of our GW emission, our model can still agree with current constraints on MG. The discussion about the constraint of $\alpha$ goes beyond the scope of this paper and requires further investigation.
Furthermore, notice that the GR solution is again recovered in the deep infrared regime \footnote{This behavior could also be seen in Fig. \ref{fig:LVK_spectrum} if the frequency range were extended.} and this can be shown analytically as demonstrated in Ref. \cite{Domenech:2024drm}. Specifically, for $x\ll1$ (deep infrared) and assuming a primordial spectrum with a sharp peak, the $f(R)$ gravity contribution is suppressed by $\mathcal{O}(\alpha x^2)$, ensuring that the GR result is always recovered.

\begin{figure}[h]
    \centering
    \includegraphics[width=0.75\textwidth]{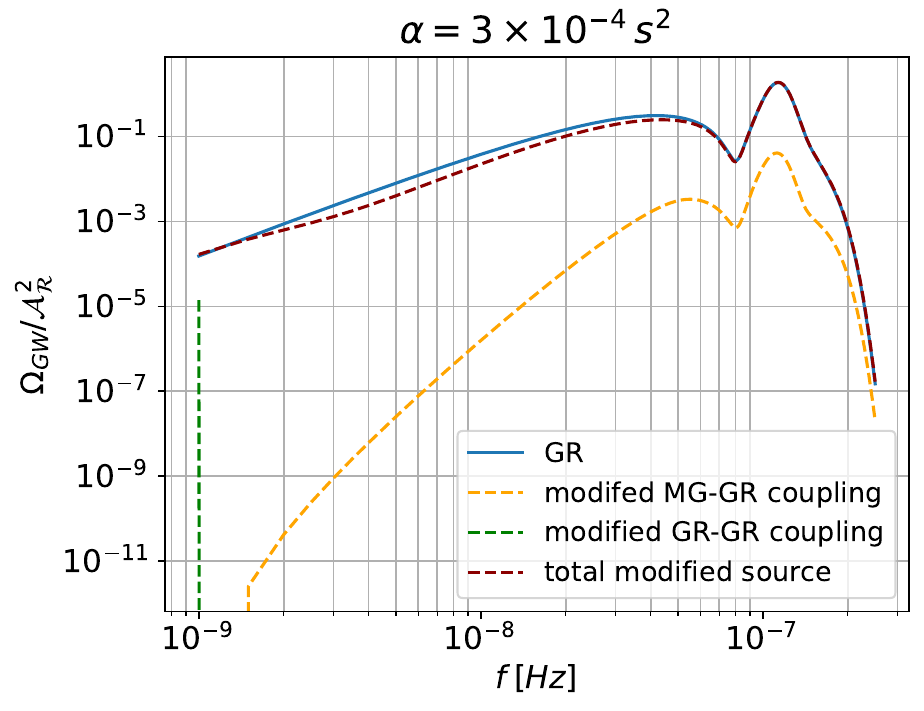}
    \caption{Energy density spectrum of gravitational waves in the PTA frequency range, showing the separate contributions from the MG-GR coupling (first four lines in Eq.\eqref{modifed_source_function_radiation}), GR-GR coupling (latter lines of Eq.\eqref{modifed_source_function_radiation}), and the total modified gravity (MG) source. Both axes are plotted on logarithmic scales to highlight features across several orders of magnitude in frequency and amplitude.}
    \label{fig:gw_energy_spectrum}
\end{figure}

To assess which component of the source term in Eq.~\eqref{modifed_source_function_radiation} plays a dominant role—the gravitational slip (MG-GR coupling, from the first four lines) or the modified source terms (GR-GR coupling, from the latter lines)—we analyze their respective impacts on the resulting SIGW signal. As shown in Fig.~\ref{fig:gw_energy_spectrum}, the MG-GR coupling consistently exceeds the GR-GR contribution across most of the PTA frequency range. This is confirmed by a pointwise analysis\footnote{A pointwise analysis compares the contributions of different terms at each evaluation point rather than using integrated quantities.}, which shows that the MG-GR term dominates in 62.8\% of the sampled points. It's relative contribution is a median of 87.24\% to the total source, indicating that the MG-GR coupling is generally the leading contributor to the modified gravity SIGW signal.

\newpage

\section{\label{sec:conclusion}Conclusions and discussion}

In this work, we conduct a comprehensive analysis of second-order scalar-induced gravitational waves (SIGWs) generated by primordial curvature perturbations re-entering the horizon during the radiation-dominated era within a modified gravity framework. The source term of SIGWs is proportional to the gravitational slip, which characterizes the difference between the first-order scalar metric potentials. In general relativity (GR), this term corresponds to the anisotropic stress induced by free-streaming neutrinos and is typically negligible. However, in extensions of GR, modifications to the geometric structure of Einstein's equations can lead to a non-negligible gravitational slip, altering the SIGW source term and potentially resulting in significant contributions to the second-order gravitational waves.

We explore the relevance of these effects within a specific $f(R)$ gravity model, with $f(R) = R + \alpha R^2$. We compute the corrections to the first- and second-order perturbations at leading order in the coupling parameter $\alpha$. We then derive a general expression for the kernel and the power spectrum and, using a log-normal primordial spectrum for the curvature fluctuations, we determine the SIGWs spectral energy density $\Omega_{\text{GW}} =\Omega_{\text{GR,GW}}+2\alpha\delta \Omega_{\text{GW}}$, assessing the impact of beyond-GR corrections. We perform a semi-numerical analysis on the SIGW energy density spectrum in various frequency ranges that will be spanned by future GW detectors like PTA and LISA, and by current ground-based interferometers. 

We notice that the $f(R)$ contribution has the same qualitative behavior of the GR background and does not add any large distinctive features to the SIGWs spectrum. The only impact is visible in the low frequency tail of the spectrum. This might be due to the choice of model and the bound on $\alpha$. We impose constraints on $\alpha$ to ensure that the system remains within the perturbative regime and that the GW energy density remains positive, requiring the beyond-GR contribution to be subdominant compared to the GR contribution. Notably, constraints on $\alpha$ derived from cosmic microwave background (CMB) observations are not applicable in this context, as the considered $f(R)$ model is not related to the Starobinsky model during inflation. A more detailed quantitative analysis of the parameter space relevant to various GW detectors will be presented in a forthcoming study.

Lastly, we note that our formalism can be easily extended to a broader class of scalar-tensor theories, which will be considered in a follow-up study. 

\acknowledgments
 We thank N. Bartolo, D. Bertacca, G. Perna for useful comments and discussions. A.M. acknowledges financial support from MUR PRIN Grant No. 2022-Z9X4XS, funded by the European Union - Next Generation EU. S.M. acknowledges financial support from the COSMOS network (www.cosmosnet.it) through the ASI (Italian Space Agency) Grants 2016-24-H.0, 2016-24-H.1-2018 and 2020-9-HH.0.

\newpage

\appendix

\section{\label{sec:appendixB}Modified background equations}

Beyond-GR corrections to the background equations are computed from the time-time and spatial components of Eq. ~\eqref{field_equation_2}, 
\begin{equation}
\label{generic_modified_background_00}
    a^{2}\kappa^{2}\bar{\rho} = 3\mathcal{H}^{2}\bar{F} + 3\mathcal{H}\bar{F}' - \frac{1}{2}a^{2}(\bar{F}R^{(0)}-\bar{f}) \, ,
\end{equation}
\begin{equation}
\label{generic_modified_background_ij}
    2(\mathcal{H}^{2} - \mathcal{H}')\bar{F} =a^{2}\kappa^{2}(\bar{\rho} + \bar{P}) + \bar{F}'' - 2\mathcal{H}\bar{F}'\,.
\end{equation}
In the $f(R)$ model we consider, the above modified equations reduce to 
\begin{equation}
\label{modified_background_00}
    a^{2}\kappa^{2}\bar{\rho} = 3\mathcal{H}^{2} + 3a^{-2}\alpha\left[12\mathcal{H}''\mathcal{H}- 18\mathcal{H}^{4} - 6(\mathcal{H}')^{2}\right],
\end{equation}
\begin{equation}
\label{modified_background_ij}
    \frac{a^{2}\kappa^{2}}{3}\left(\bar{\rho}+3\bar{P}\right) = -2\mathcal{H}' + 2a^{-2}\alpha\left[12\mathcal{H}''\mathcal{H} - 18\mathcal{H}^{4} - 6(\mathcal{H}')^{2} + 36\mathcal{H}'\mathcal{H}^{2} - 6\mathcal{H}'''\right].
\end{equation}
We also introduce beyond-GR changes to the Hubble parameter and the scale factor as  $\mathcal{H} = \mathcal{H}_{\text{GR}} + \alpha\delta\mathcal{H}$ and $a = a_{\text{GR}} + \alpha\delta a$, where $\delta\mathcal{H}$ can be written in terms of $a$ as
\begin{align*}
\label{modified_hubble}
    \mathcal{H}_{\text{GR}} + \alpha\delta\mathcal{H} & = \frac{(a_{\text{GR}} + \alpha\delta a)'}{a_{\text{GR}} + \alpha\delta a}, \\
    \alpha\delta\mathcal{H} & \simeq \frac{\alpha}{a_{\text{GR}}}\left(\delta a' - \mathcal{H}_{\text{GR}}\delta a\right). \numberthis
\end{align*}
Using these definitions we expand Eqs.~\eqref{modified_background_00}-\eqref{modified_background_ij}, with terms proportional to $\alpha$ yielding
\begin{equation}
\label{expanded_mg_00}
    \kappa^{2}\left[2 a_{\text{GR}} \delta a \rho_{\text{GR}}^{(0)} + a_{\text{GR}}^{2}\delta\bar{\rho}\right] = 6\mathcal{H}_{\text{GR}}\delta\mathcal{H} + 3a^{-2}\left[12\mathcal{H}_{\text{GR}}''\mathcal{H}_{\text{GR}}- 18\mathcal{H}_{\text{GR}}^{4} - 6(\mathcal{H}_{\text{GR}}')^{2}\right],
\end{equation}
\begin{align*}
\label{expanded_mg_ij}
    &\frac{\kappa^{2}(1+3w)}{3}\left[2 a_{\text{GR}}\, \delta a\, \rho_{\text{GR}}^{(o)} + a^{2}_{\text{GR}}\delta\bar{\rho}\right]=\\   &=-2\delta\mathcal{H}' + 2a^{-2}\left[12\mathcal{H}_{\text{GR}}''\mathcal{H}_{\text{GR}} - 18\mathcal{H}_{\text{GR}}^{4} - 12(\mathcal{H}_{\text{GR}}')^{2} + 36\mathcal{H}_{\text{GR}}'\mathcal{H}_{\text{GR}}^{2} - 6\mathcal{H}_{\text{GR}}'''\right].\\ \numberthis
\end{align*}
In the radiation-dominated epoch, and assuming $w=1/3$, $a_{\text{GR}} \propto \tau$ and $\mathcal{H}_{\text{GR}} = \tau^{-1}$, the background equations (\ref{expanded_mg_00}) and (\ref{expanded_mg_ij}) simplify to
\begin{equation}
\label{background_00_rad_mg}
    \kappa^{2}\left[2 \tau^{-3}\delta a + \tau^{2}\delta\bar{\rho}\right] = 6\tau^{-1}\delta\mathcal{H}\ ,  \numberthis
\end{equation}
\begin{equation}
\label{background_ij_rad_mg}
    \frac{2\kappa^{2}}{3}\left[2\tau^{-3}\delta a + \tau^{2}\delta\bar{\rho}\right] = -2\delta\mathcal{H}'\ ,\numberthis
\end{equation}
Moreover, we use the continuity equation, $\bar{\rho}' + 3\mathcal{H}(\bar{\rho}+\bar{P}) = 0$, to find the solution for $\bar{\rho}_{\text{GR}} \propto \tau^{-4}$. By substituting Eq.~\eqref{background_00_rad_mg} into Eq.~\eqref{background_ij_rad_mg}, we can solve for $\delta a$, namely we get
\begin{equation}
\label{solution_delta_a}
    \delta{a} = c_{1}\tau + c_{2}\ ,
\end{equation}
and using Eq.~\eqref{modified_hubble} we find $\delta\mathcal{H}$ as
\begin{equation}
\label{solution_delta_H} 
    \delta\mathcal{H} = \frac{c_{2}}{\tau^{2}}\ . 
\end{equation}
The behavior of the modified Hubble parameter decays faster than its GR counterpart for large $\tau$, and can be neglected.

\section{\label{sec:appendixC}Beyond-GR source term}

In Eq. (\ref{second_order_tensor_equation_mg}), the full source term in $f(R)$ gravity is
\begin{align*}
\label{second_order_source_term_modified}
S_{j}^{i} + \alpha\delta S_{j}^{i} &= - \left(1 + 12a^{-2}\alpha\left[\mathcal{H}' + \mathcal{H}^{2}\right]\right)\left(\partial^{i}\firstphi\partial_{j}\firstphi + 2\firstphi\partial^{i}\partial_{j}\firstphi \right. \\
    & \left. - 2\firstpsi\partial^{i}\partial_{j}\firstphi - \partial_{j}\firstphi\partial^{i}\firstpsi - \partial^{i}\firstphi\partial_{j}\firstpsi \right. \\
    & \left. + 3\partial^{i}\firstpsi\partial_{j}\firstpsi + 4\firstpsi\partial^{i}\partial_{j}\firstpsi\right) \\
    & - 2a^{-2}\alpha\left( - 6\ddfirstpsi - 6\mathcal{H}\dfirstphi  - 18\mathcal{H}\dfirstpsi  - 12\left[\mathcal{H}' + \mathcal{H}^{2}\right]\firstphi \right. \\
    &\left.  - 2\nabla^{2}\firstphi + 4\nabla^{2}\firstpsi\right)\left(\partial^{i}\partial_{j}\firstpsi - \partial^{i}\partial_{j}\firstphi\right) \\
    & + 4a^{-2}\alpha\firstpsi\partial^{i}\partial_{j}\left( - 6\ddfirstpsi - 6\mathcal{H}\dfirstphi - 18\mathcal{H}\dfirstpsi \right. \\
    & \left. - 12\left[\mathcal{H}' + \mathcal{H}^{2}\right]\firstphi - 2\nabla^{2}\firstphi + 4\nabla^{2}\firstpsi\right) \\
    & + 2a^{-2}\alpha\partial^{i}\partial_{j}\left(24\left[\mathcal{H}' + \mathcal{H}^{2}\right](\firstphi)^{2} + 24\mathcal{H}\firstphi\dfirstphi + 6\dfirstphi\dfirstpsi \right. \\
    &\left.  + 36\mathcal{H}\firstphi\dfirstpsi + 12\firstphi\ddfirstpsi - 12\firstpsi\ddfirstpsi - 36\mathcal{H}\firstpsi\dfirstpsi  + 4\firstphi\nabla^{2}\firstphi \right. \\
    &\left.  - 4\firstpsi\nabla^{2}\firstphi + 16\firstpsi\nabla^{2}\firstpsi + 2\partial^{k}\firstphi\partial_{k}\firstphi \right. \\
    &\left.  + 2\partial^{k}\firstphi\partial_{k}\firstpsi + 6\partial^{k}\firstpsi\partial_{k}\firstpsi\right) \\
    & + 2a^{-2}\alpha\partial^{i}\firstpsi\partial_{j}\left( - 6\ddfirstpsi - 6\mathcal{H}\dfirstphi - 18\mathcal{H}\dfirstpsi \right. \\
    & \left. - 12\left[\mathcal{H}' + \mathcal{H}^{2}\right]\firstphi - 2\nabla^{2}\firstphi + 4\nabla^{2}\firstpsi\right)\\
    & + 2a^{-2}\alpha\partial_{j}\firstpsi\partial^{i}\left( - 6\ddfirstpsi - 6\mathcal{H}\dfirstphi - 18\mathcal{H}\dfirstpsi \right. \\
    & \left. - 12\left[\mathcal{H}' + \mathcal{H}^{2}\right]\firstphi - 2\nabla^{2}\firstphi + 4\nabla^{2}\firstpsi\right) \\
    & + \frac{4}{3\mathcal{H}^{2}(1+w)}\left[\partial^{i}\left(\dfirstpsi + \mathcal{H}\firstphi\right)\partial_{j}\left(\dfirstpsi + \mathcal{H}\firstphi\right)\right] \\
    & - \frac{4a^{-2}\alpha\left[12\mathcal{H}''\mathcal{H}- 18\mathcal{H}^{4} - 6(\mathcal{H}')^{2}\right]}{3\mathcal{H}^{4}(1+w)}\left[\partial^{i}\left(\dfirstpsi + \mathcal{H}\firstphi\right)\partial_{j}\left(\dfirstpsi + \mathcal{H}\firstphi\right)\right] \\
    & + \frac{2\alpha a^{-2}}{3\mathcal{H}^{2}(1+w)}\partial^{i}\left(\dfirstpsi + \mathcal{H}\firstphi\right)\partial_{j}\left([36\mathcal{H}'' - 72\mathcal{H}^{3}]\firstphi + [36\mathcal{H}' - 12\mathcal{H}^{2}]\dfirstphi \right. \\
    & \left. + [60\mathcal{H}' - 84\mathcal{H}^{2}]\dfirstpsi + 12\mathcal{H}\ddfirstphi + 12\dddfirstpsi \right. \\
    & \left. + 12\mathcal{H}\left(2\nabla^{2}\firstpsi - \nabla^{2}\firstphi\right) + 4\left(\nabla^{2}\dfirstphi - 2\nabla^{2}\dfirstpsi\right)\right)\\
    & + \frac{2\alpha a^{-2}}{3\mathcal{H}^{2}(1+w)}\partial_{j}\left(\dfirstpsi + \mathcal{H}\firstphi\right)\partial^{i}\left([36\mathcal{H}'' - 72\mathcal{H}^{3}]\firstphi + [36\mathcal{H}' - 12\mathcal{H}^{2}]\dfirstphi \right. \\
    & \left. + [60\mathcal{H}' - 84\mathcal{H}^{2}]\dfirstpsi + 12\mathcal{H}\ddfirstphi + 12\dddfirstpsi \right. \\
    & \left. + 12\mathcal{H}\left(2\nabla^{2}\firstpsi - \nabla^{2}\firstphi\right) + 4\left(\nabla^{2}\dfirstphi - 2\nabla^{2}\dfirstpsi\right)\right) \\
    & - 4\firstpsi \delta^{ik}\left(1 + 12\alpha a^{-2}\left[\mathcal{H}' + \mathcal{H}^{2}\right]\right)\left(\partial_{j}\partial_{k} - \frac{1}{3}\nabla^{2}\delta_{jk}\right)\left(\firstphi - \firstpsi\right) \\
    & - 8\alpha a^{-2}\firstpsi \delta^{ik}\left(\partial_{j}\partial_{k} - \frac{1}{3}\nabla^{2}\delta_{jk}\right)\left( - 6\ddfirstpsi - 6\mathcal{H}\dfirstphi - 18\mathcal{H}\dfirstpsi \right. \\
    & \left. - 12\left[\mathcal{H}' + \mathcal{H}^{2}\right]\firstphi  - 2\nabla^{2}\firstphi + 4\nabla^{2}\firstpsi\right). \numberthis
\end{align*}
As we are only considering the first-order $\alpha$ correction, we expand the source term~\eqref{second_order_source_term_modified} to first-order in $\alpha$ using Eqs.~\eqref{modified_phi_def} -~\eqref{modified_p_def} and Eq.~\eqref{delta_psi_definition}. By doing so, we find the following expression
\begin{align*}
\label{second_order_source_term_modified_expanded}
    \delta S_{j}^{i} &= -2\partial^{i}\delta\firstphi\partial_{j}\firstphigr - 2\partial^{i}\firstphigr\partial_{j}\delta\firstphi - 4\delta\firstphi\partial^{i}\partial_{j}\firstphigr - 4\firstphigr\partial^{i}\partial_{j}\delta\firstphi\\
    & + \frac{4}{3\mathcal{H}^{2}(1+w)}\left[\partial^{i}\delta\dfirstphi\partial_{j}\dfirstphigr + \partial_{j}\delta\dfirstphi\partial^{i}\dfirstphigr\right] \\
    & + \frac{4}{3\mathcal{H}(1+w)}\left[\partial^{i}\delta\dfirstphi\partial_{j}\firstphigr + \partial_{j}\delta\dfirstphi\partial^{i}\firstphigr + \partial^{i}\delta\firstphi\partial_{j}\dfirstphigr + \partial_{j}\delta\firstphi\partial^{i}\dfirstphigr\right] \\
    & + \frac{4}{3(1+w)}\left[\partial^{i}\delta\firstphi\partial_{j}\firstphigr + \partial_{j}\delta\firstphi\partial^{i}\firstphigr\right] \\
    & - 12a^{-2}\left[\mathcal{H}' + \mathcal{H}^{2}\right]\left(2\partial^{i}\firstphigr\partial_{j}\firstphigr + 4\firstphigr\partial^{i}\partial_{j}\firstphigr\right)\\
    &\left.  + 16\firstphigr\nabla^{2}\firstphigr + 10\partial^{k}\firstphigr\partial_{k}\firstphigr\right) \\
    & - 2a^{-2}\partial^{i}\firstphigr\partial_{j}\left( - 6\ddfirstphigr - 24\mathcal{H}\dfirstphigr - 12\left[\mathcal{H}' + \mathcal{H}^{2}\right]\firstphigr + 2\nabla^{2}\firstphigr\right)\\
    & - 2a^{-2}\partial_{j}\firstphigr\partial^{i}\left( - 6\ddfirstphigr - 24\mathcal{H}\dfirstphigr - 12\left[\mathcal{H}' + \mathcal{H}^{2}\right]\firstphigr + 2\nabla^{2}\firstphigr\right) \\
    & - 4a^{-2}\firstphigr\partial^{i}\partial_{j}\left( - 6\ddfirstphigr - 24\mathcal{H}\dfirstphigr - 12\left[\mathcal{H}' + \mathcal{H}^{2}\right]\firstphigr + 2\nabla^{2}\firstphigr\right) \\
    & - 4a^{-2}\left( - 6\ddfirstphigr - 24\mathcal{H}\dfirstphigr - 12\left[\mathcal{H}' + \mathcal{H}^{2}\right]\firstphigr + 2\nabla^{2}\firstphigr\right)\partial^{i}\partial_{j}\firstphigr \\
    & - \frac{4a^{-2}\left[12\mathcal{H}''\mathcal{H}- 18\mathcal{H}^{4} - 6(\mathcal{H}')^{2}\right]}{3\mathcal{H}^{4}(1+w)}\left[\partial^{i}\left(\dfirstphigr + \mathcal{H}\firstphigr\right)\partial_{j}\left(\dfirstphigr + \mathcal{H}\firstphigr\right)\right] \\
    & + 2a^{-2}\partial^{i}\partial_{j}\left(24\left[\mathcal{H}' + \mathcal{H}^{2}\right](\firstphigr)^{2} + 24\mathcal{H}\firstphigr\dfirstphigr + 6\dfirstphigr\dfirstphigr \right. \\
    & \left. + 16\firstphigr\nabla^{2}\firstphigr + 10\partial^{k}\firstphigr\partial_{k}\firstphigr \right) \\
    & + \frac{2 a^{-2}}{3\mathcal{H}^{2}(1+w)}\partial^{i}\left(\dfirstphigr + \mathcal{H}\firstphigr\right)\partial_{j}\left([36\mathcal{H}'' - 72\mathcal{H}^{3}]\firstphigr + [96\mathcal{H}' - 96\mathcal{H}^{2}]\dfirstphigr  \right. \\
    & \left. + 12\mathcal{H}\ddfirstphigr + 12\dddfirstphigr + 12\mathcal{H}\nabla^{2}\firstphigr -4\nabla^{2}\dfirstphigr\right)\\
    & + \frac{2 a^{-2}}{3\mathcal{H}^{2}(1+w)}\partial_{j}\left(\dfirstphigr + \mathcal{H}\firstphigr\right)\partial^{i}\left([36\mathcal{H}'' - 72\mathcal{H}^{3}]\firstphigr + [96\mathcal{H}' - 96\mathcal{H}^{2}]\dfirstphigr  \right. \\
    & \left. + 12\mathcal{H}\ddfirstphigr + 12\dddfirstphigr + 12\mathcal{H}\nabla^{2}\firstphigr -4\nabla^{2}\dfirstphigr\right) \\
    & + \frac{8 a^{-2}}{3\mathcal{H}^{2}(1+w)}\partial^{i}\left(\dfirstphigr + \mathcal{H}\firstphigr\right)\partial_{j}\left(-[12\mathcal{H}'' - 24\mathcal{H}^{3}]\firstphigr - [36\mathcal{H}' - 36\mathcal{H}^{2}]\dfirstphigr  \right. \\
    & \left. -12\mathcal{H}\ddfirstphigr - 6\dddfirstphigr - 4\mathcal{H}\nabla^{2}\firstphigr + 2\nabla^{2}\dfirstphigr\right)\\
    & + \frac{8 a^{-2}}{3\mathcal{H}^{2}(1+w)}\partial_{j}\left(\dfirstphigr + \mathcal{H}\firstphigr\right)\partial^{i}\left(-[12\mathcal{H}'' - 24\mathcal{H}^{3}]\firstphigr - [36\mathcal{H}' - 36\mathcal{H}^{2}]\dfirstphigr  \right. \\
    & \left. -12\mathcal{H}\ddfirstphigr - 6\dddfirstphigr - 4\mathcal{H}\nabla^{2}\firstphigr + 2\nabla^{2}\dfirstphigr\right) \ . \numberthis 
\end{align*}
%


\newpage

\bibliography{letter_SIGW_modified_gravityNotes.bib}

%\begin{thebibliography}{99}

%\bibitem{a}
%Author, \emph{Title}, \emph{J. Abbrev.} {\bf vol} (year) pg.

%\bibitem{b}
%Author, \emph{Title},
%arxiv:1234.5678.

%\bibitem{c}
%Author, \emph{Title},
%Publisher (year).

%\end{thebibliography}

\end{document}